\documentclass[useAMS,usenatbib]{mn2e}

\usepackage{psfig, epsf, epsfig}

\title[IMF-regulated galaxy evolution]{Simulating galaxy evolution
with a non-universal stellar initial mass function}
\author[K. Bekki]
{Kenji Bekki${}^1$\thanks{E-mail:
bekki@cyllene.uwa.edu.au} \\
${}^1$ICRAR M468
The University of Western Australia
35 Stirling Hwy, Crawley
Western Australia 6009, Australia}

\begin{document}

\date{Accepted, Received 2005 February 20; in original form }

\pagerange{\pageref{firstpage}--\pageref{lastpage}} \pubyear{2005}

\maketitle

\label{firstpage}

\begin{abstract}
We consider that the stellar initial mass function (IMF) depends on physical
properties of star-forming molecular clouds in galaxies and thereby investigate
how such a non-universal IMF (NUIMF) influences  galaxy  evolution.
We incorporate a NUIMF model into galaxy-scale chemodynamical simulations 
in order to
investigate the differences in chemical and dynamical evolution of disk 
galaxies between the NUIMF and universal IMF (UIMF) models.
In the adopted NUIMF model, the three slopes 
of the Kroupa IMF depend
independently on densities and metallicities ([Fe/H]) of molecular gas clouds,
and production rates of metals and dust from massive 
and AGB stars,
formation efficiencies
of molecular  hydrogen (${\rm H}_2$),
and feedback effects of supernovae (SNe)
can vary according to the time  evolution of the three IMF slopes.
The preliminary results of the simulations are as follows.
Star formation rates (SFRs) in actively star-forming
disk galaxies can be significantly lower in the  
NUIMF model than in the UIMF model, and the differences 
between the two models
can be larger  in galaxies with higher SFRs.
Chemical enrichment can proceed faster in the NUIMF model 
and [Mg/Fe] for a given metallicity is higher
in the NUIMF model. 
The evolution of ${\rm H_2}$ fraction ($f_{\rm H_2}$)
and dust-to-gas ratio ($D$) 
is faster in the NUIMF model
so that the final $f_{\rm H_2}$ and  $D$ can be higher in the NUIMF model.
Formation of massive stellar clumps in gas-rich disks is 
more strongly suppressed owing to the stronger SN feedback effect
in the NUIMF model.
The radial density profiles of new stars within the central 1kpc are shallower
in the NUIMF model.
\end{abstract}


\section{Introduction}

The stellar initial mass function (IMF) is a fundamentally important
factor for galaxy formation and evolution,  because it can control
the number evolution of stars with different masses in galaxies
and thus the  long-term evolution
of luminosities, chemical abundances, dynamical properties 
of galaxies (e.g., Larson 1998).
Physical interpretation of observational data and theoretical predictions
of galactic properties therefore depend strongly on what IMF is
adopted. It is still controversial, however, whether the IMF is universal
or different in galaxies with different physical properties
(e.g., Bastian et al. 2010; Kroupa et al. 2012).
It is also observationally unclear whether or how the IMF depends on local physical
properties of star-forming giant molecular clouds in galaxies with different
properties (e.g., masses and luminosities),
though previous  theoretical studies of star formation investigated
possible dependences of the IMF slopes on cloud properties
(e.g., Elmegreen 2004; Larson 2005).

A growing number of recent observational studies have found possible
evidences for (i) a significant variation of the IMF
in galaxies with different physical properties and different redshifts
and (ii) correlations between the IMF slope and galactic properties
(e.g.,  Hoversten \& Glazebrook 2008;
van Dokkum 2008; Meurer et al. 2009, M09; Treu et al. 2010;
Gunawardhana et al. 2011, G11;
Cappellari et al. 2012; Conroy \& van Dokkum 2012;
Ferreras et al. 2012; van Dokkum \& Conroy 2012; Geha et al. 2013).
These observational results have raised a number of questions on
the IMF such as (i) how the IMF can be bottom-heavy in the formation
of elliptical galaxies (e.g., van Dokkum \& Conroy 2012),
(ii) how the IMF slope depends on star formation rates and densities
in galaxies (e.g., M09; G11), and 
(iii) whether or not  galaxy formation models with a universal
IMF (UIMF) can explain the observed 
Faber-Jackson and Tully-Fisher relation of galaxies
(e.g., Dutton et al. 2011).

A number of previous theoretical  studies adopted a non-standard 
IMF to explain 
some specific observational results on physical properties of 
galaxies
(e.g., Elmegreen 2009 for a recent review).
Their models with top-heavy IMFs 
tried to explain the observed colors of elliptical
galaxies (e.g., Pipino \& Matteuci 2004),
the chemical abundances (e.g., [$\alpha$/Fe]) of giant elliptical
galaxies (e.g., Nagashima et al. 2005),
and the number counts of
distant  submillimeter galaxies (Baugh et al. 2005).
These studies chose a fixed non-standard IMF either for the entire galaxy
formation or only for starburst phases in  an arbitrary manner
in order to  explain some specific 
observational results.  
These studies are therefore unable to discuss how the IMF {\it should}  change
in different galaxies and in different environments 
in order to explain different observational results 
{\it in a self-consistent manner.}

So far, only a few theoretical studies have investigated the time evolution
of the IMF in galaxies with different physical
properties based on hydrodynamical simulations of galaxy
formation and evolution with a NUIMF model (e.g., Dav\'e 2008;
Narayana \& Dav\'e 2012;
Bekki \& Meurer 2013, BM13). These simulations, however, did not consider
the possible physical effects of a NUIMF on galaxies
thus were unable to discuss whether and how a NUIMF
can influence the formation and evolution processes of galaxies in 
a self-consistent manner.
Although the number of SNe, chemical yields, and dust yields
are significantly different between different IMFs,
previous  numerical simulations of galaxy formation and evolution with a NUIMF
model did not  include 
the dependences
of SN feedback effects,
chemical evolution, and dust production processes on the IMF.
Therefore, it is essential for theoretical studies to include self-consistently
the IMF-dependent physical processes in order to discuss
how galaxy evolution is influenced by a NUIMF.

Thus the two major purposes of this paper are  as follows. One is to describe 
the details of our new chemodynamical simulations  with a NUIMF
model. The other is to show the preliminary results of the simulations 
on the major influences of the adopted NUIMF on SFRs, chemical and dynamical
evolution, and dust and ${\rm H_2}$ contents of disk galaxies.
We implement the NUIMF model recently proposed by Marks et al. (2012, M12)
in the present study.
This is mainly
because the NUIMF model by M12  was derived from comparison
between observational data sets and theoretical modeling.
M12 showed that (i) the stellar 
present-day mass function (PDMF) in the Galactic globular clusters
(GCs) can be determined  by residual-gas expulsion at the epoch of GC formation,
(ii) the gas expulsion process depends on the high-mass IMF ($\alpha_3$) of forming GCs,
and  (iii) the observed PDMF depending on GC properties
(half-mass radius, mass density, and metallicity) can be reproduced by  their NUIMF model
that depends on density and metallicity of forming GCs. 
Although the adopted NUIMF model has some physical basis (as above),
we briefly discuss advantages and
disadvantages of the present chemodynamical simulations with
the NUIMF model by M12 later in this paper.

\begin{table}
\centering
\begin{minipage}{80mm}
\caption{Description of the basic parameter values
for the fiducial model.}
\begin{tabular}{cc}
{Physical properties}
& {Parameter values}\\
{Total Mass \footnote{$M_{\rm gal}=M_{\rm dm}+M_{\rm d}+M_{\rm b}$, where
$M_{\rm dm}$, $M_{\rm d}$, and $M_{\rm b}$
are the total masses of dark matter halo,
disk, and bulge in a galaxy, respectively.}}
& $M_{\rm gal}=1.08 \times 10^{12} {\rm M}_{\odot}$  \\
{DM structure \footnote{ The NFW profile with a virial radius ($R_{\rm vir}$)
and a $c$ parameter is adopted for the structure of dark matter halo.}}
& $R_{\rm vir}=120$ kpc,  $c=10$  \\
Disk mass & $M_{\rm d}=6.6 \times 10^{11} {\rm M}_{\odot}$     \\
Disk scale length & $R_{0, \rm s}=3.5$ kpc \\
Gas fraction in a disk & $f_{\rm g}=0.09$     \\
Bulge mass &   $M_{\rm b}=10^{10} {\rm M}_{\odot}$  \\
Bulge size  & $R_{\rm b}=3.5$ kpc \\
Galaxy interaction   &  No \\
Initial central metallicity   &   ${\rm [Fe/H]_0}=0.34$ \\
Initial  metallicity gradient   &   $\alpha_{\rm d}=-0.04$ dex kpc$^{-1}$ \\
Chemical yield  &  T95 for SN,  VG97 for AGB \\
Dust yield  &  B13 \\
{Dust formation model  \footnote{$\tau_{\rm acc}$ and $\tau_{\rm dest}$
are the dust accretion and destruction timescales, respectively.}}
& $\tau_{\rm acc}=0.25$ Gyr, $\tau_{\rm dest}=0.5$ Gyr  \\
Initial dust/metal ratio  & 0.4  \\
{${\rm H_2}$ formation  \footnote{The same model as B13, where
${\rm H_2}$ formation processes depend on local gas density, $D$ (dust-to-gas
ratio), and interstellar radiation field (ISRF). }}
& Dependent on $D$ and ISRF  (B13) \\
{Feedback \footnote{$f_{\rm b}$ is the binary fraction of stars that
can finally explode as SNe Ia. The fiducial  model without SN feedback effects
is also investigated for comparison. No AGN feedback effects are 
included in the present study.}}  & SNIa ($f_{\rm b}=0.09$) and SNII \\
{SF \footnote{$\rho_{\rm th}$ is the threshold gas density for star formation
and interstellar radiation field (ISRF) is included in the estimation of
${\rm H_2}$ mass fraction in this model.}}
& ${\rm H}_2$-dependent,  ISRF,  $\rho_{\rm th}=1$ cm$^{-3}$ \\
{IMF \footnote{The time-varying Kroupa IMF is referred to
as `NUIMF'. This standard model with the fixed 
Kroupa IMF model (`UIMF')  is also  investigated
for comparison.}}  & Time-varying Kroupa IMF \\
Particle number & $N=1033400$ \\
Softening length  & $\epsilon_{\rm dm}=2.1$ kpc, $\epsilon_{\rm s}=200$ pc \\
Mass resolution for stars   & $2.7 \times 10^5 {\rm M}_{\odot}$ \\
\end{tabular}
\end{minipage}
\end{table}

The plan of this  paper is as follows: In the next section,
we describe the methods and techniques of
our new chemodynamical simulations with a NUIMF model.
In \S 3, we clearly demonstrate the predictive power of
the new simulations of  IMF evolution and its influences
on galaxy formation and evolution.
In \S 4, we present the numerical results
on the time  evolution of physical properties of isolated
and interacting disk galaxies, such as star formation rates (SFRs),
dust-to-gas ratios ($D$), molecular fraction ($f_{\rm H_2}$),
chemical abundances, and dynamical properties.
We discuss the important implications of the present results in terms of
(i) possible evidence of top-heavy IMFs and (ii)  bulge formation
in \S 5.
We summarize our  conclusions in \S 6.

The present study  focuses exclusively on the influences of the NUIMF on the evolution
of disk galaxies.
Unlike previous theoretical studies on neutral and molecular hydrogen
and chemical abundances of forming galaxies  based on a CDM cosmology
(e.g., Fu et al. 2010, Lagos et al. 2011; Duffy et al. 2012; Dave et al. 2013),
the present simulations do not start from initial conditions of
galaxies formation.  Therefore, the present study does not allow us to discuss 
physical properties of all components (gas, metals, old and young stars) in
a fully self-consistently manner. However, the results of the present simulations can help
us to better understand the possible influences of the NUIMF on galaxy evolution.
Our future simulations of galaxy formation with the NUIMF will address key issues 
of galaxy formation discussed in the above previous theoretical studies.

\section{The model}

\subsection{Overview}

We perform numerical simulations of
actively star-forming disk galaxies by using
our original chemodynamical code (Bekki 2013, B13) which
incorporated formation and evolution of ${\rm H_2}$
and accretion and destruction of dust in a self-consistent manner.
In the present study, we newly incorporate a NUIMF
into the chemodynamical simulations so that we can discuss
both (i) the time evolution of the IMF and (ii) its influences on
the evolution of chemical and dynamical properties and dust and ${\rm H_2}$
contents in galaxies. Since the details of the chemodynamical code
is given in B13, we just briefly describe the code and focus on
how we implement a NUIMF in galaxy-scale chemodynamical 
simulations in the present study.

The three IMF slopes of the Kroupa IMF for each star-forming gas cloud in
a galaxy are assumed to depend on physical properties of the clouds
and be therefore time-varying in the present study. We estimate
the IMF slopes for each cloud at each time step in a chemodynamical
simulation by examining the cloud properties. The chemical yields and dust
production from SNe and AGB stars strongly depend on the IMF slopes and
need to be self-consistently changed according to the slopes. We 
therefore include the IMF-dependent chemical and dust yields in the 
present chemodynamical evolution.  Also SN feedback effects,  ${\rm H_2}$
formation on dust grains, and dust accretion and destruction, all of which
can depend on the IMF slopes, are self-consistently included in the
simulations according to the changes of the IMF slopes.

\begin{table*}
\centering
\begin{minipage}{175mm}
\caption{Parameter values for the representative models.}
\begin{tabular}{ccccccccc}
{Model name \footnote{ For each model (e.g. M1), galaxy evolution
with the UIMF or the NUIMF is investigated.
Also, for each model, 
galaxy evolution with the NUIMF and with/without SN feedback
effects (`SN' or `NSN', respectively) is investigated.}}
& { $M_{\rm dm}$ (${\rm M}_{\odot}$) }
& { $c$ }
& { $f_{\rm dm}$ }
& { $f_{\rm b}$ }
& { $f_{\rm g}$ }
& { [Fe/H]$_0$ }
& { $m_2$ 
\footnote{The mass-ratio of two interacting galaxies is represented
by $m_2$. If tidal interaction with a companion galaxy is not included,
$m_2=0$ is shown.}}
& Comments \\
M1 & $10^{12}$ & 10.0 & 15.2 & 0.15 & 0.09 & 0.34 & 0.0 & isolated MW, 
gas-poor, standard  \\
M2 & $10^{12}$ & 10.0 & 15.2 & 0.15 & 0.27 & 0.17 & 0.0 &  \\
M3 & $10^{12}$ & 10.0 & 15.2 & 0.15 & 0.55 & $-0.11$ & 0.0 & gas-rich \\
M4 & $10^{12}$ & 10.0 & 15.2 & 0.15 & 0.09 & 0.34 & 1.0 & tidal interaction \\
M5 & $10^{11}$ & 12.0 & 15.2 & 0.0 & 0.55 & $-0.36$ & 0.0 & gas-rich LMC \\
M6 & $10^{10}$ & 16.0 & 15.2 & 0.0 & 0.55 & $-0.61$ & 0.0 & gas-rich dwarf \\
M7 & $10^{12}$ & 10.0 & 15.2 & 0.0 & 0.09 & 0.34 & 0.0 & bulgeless \\
M8 & $10^{12}$ & 10.0 & 15.2 & 1.0 & 0.09 & 0.34 & 0.0 & big bulge \\
M9 & $10^{12}$ & 10.0 & 30.4 & 0.30 & 0.55 & $-0.19$ & 0.0 & 
gas-rich, low-mass disk \\
M10 & $10^{12}$ & 10.0 & 75.9 & 0.76 & 0.55 & $-0.29$ & 0.0 & 
early MW disk \\
M11 & $10^{12}$ & 10.0 & 15.2 & 0.15 & 0.09 &  0.34 & 0.0 &  
LSB, $R_{\rm d,s}=26.3$ kpc\\
M12 & $10^{11}$ & 12.0 & 15.2 & 0.0 & 0.27 & $0.01$ & 0.0 &  less gas-rich LMC \\
M13 & $10^{10}$ & 16.0 & 15.2 & 0.0 & 0.27 & $-0.32$ & 0.0 &  less gas-rich Dwarf \\
M14 & $10^{12}$ & 10.0 & 15.2 & 0.15 & 0.09 &  0.34 & 3.0 &  
high-mass companion\\
M15 & $10^{12}$ & 10.0 & 15.2 & 0.15 & 0.09 &  0.34 & 0.3  & 
low-mass companion \\
\end{tabular}
\end{minipage}
\end{table*}

\subsection{Isolated spiral galaxy}

A  spiral galaxy  is composed of  dark matter halo,
stellar disk,  stellar bulge, and  gaseous disk.
Gaseous halos that were included in our previous works 
(e.g., Bekki 2009) are not included in the present paper, 
because we do not discuss
ram pressure stripping of halo gas within clusters and groups of galaxies.
The total masses of dark matter halo, stellar disk, gas disk, and
bulge are denoted as $M_{\rm dm}$, $M_{\rm d,s}$, $M_{\rm d,g}$,
and $M_{\rm b}$, respectively. The total disk mass (gas + stars)
and gas mass fraction
are denoted as $M_{\rm d}$ and $f_{\rm g}$, respectively,  for convenience.
The mass ratio of the dark matter halo to the  disk
in a disk galaxy  is a free parameter 
($f_{\rm  dm}=M_{\rm dm}/M_{\rm d}$)
and the density distribution of the halo is represented by the NFW profile
(Navarro et al. 1996). 
The $c$-parameter and the virial radius ($R_{\rm vir}$) are chosen appropriately
for a given dark halo mass ($M_{\rm dm}$) 
by using the $c-M_{\rm dm}$ relation 
predicted by recent cosmological simulations (Neto et al. 2007).

The bulge of a spiral has a size of $R_{\rm b}$
and a scale-length of $R_{\rm 0, b}$
and is represented by the Hernquist
density profile. The bulge is assumed to have isotropic velocity dispersion
and the radial velocity dispersion is given according to the Jeans equation
for a spherical system.
The bulge-mass fraction ($f_{\rm b}=M_{\rm b}/M_{\rm d}$) is
a free parameter.
We mainly investigate ``Milky Way'' models (referred to as ``MW''
from now on) in which $f_{\rm b}=0.15$ and $R_{\rm b}=0.2R_{\rm d,s}$
(i.e.,  $R_{\rm 0,b}=0.04R_{\rm d,s}$), where $R_{\rm d,s}$, is the size
of the stellar disk.
We describe  the results of  the models with $f_{\rm b}=0$ (`bulgeless'),
0.167 (MW bulge), and 1 (`big bulge') in the present study.

The radial ($R$) and vertical ($Z$) density profiles of the stellar disk are
assumed to be proportional to $\exp (-R/R_{0,s}) $ with scale
length $R_{0,s} = 0.2R_{\rm d,s}$  and to ${\rm sech}^2 (Z/Z_{0,s})$ with scale
length $Z_{0,s} = 0.04R_{\rm d,s}$, respectively.
The gas disk with a size of $R_{\rm d,g}$
has the same radial and vertical scale lengths as the 
stellar one yet has a larger size than the stellar one 
($R_{\rm d,g}=2R_{\rm d,s}$).
In the present model for the MW,  the exponential disk
has $R_{\rm d,s}=17.5$ kpc and
$R_{\rm d,g}=35$ kpc.  The adopted $R_{\rm d,g}=35$ kpc is consistent with
the recent observational result by Kalberla \& Kerp (2009) which shows
that the gas disk of the MW is well described by an exponential disk up to 35 kpc
(See Fig. 5 in their paper).
In addition to the
rotational velocity caused by the gravitational field of disk,
bulge, and dark halo components, the initial radial and azimuthal
velocity dispersions are assigned to the disc component according to
the epicyclic theory with Toomre's parameter $Q$ = 1.5.
Both gas and stellar disks in a disk galaxy have $Q=1.5$
throughout the disk  so that the disk 
can be stabilized against axisymmetric gravitational instability. However,
the adopted value of $Q=1.5$ is not enough to stabilize the disk against
global bar instability in the central regions,
 in particular, for disk galaxies with small bulges.
Therefore, the present models with small bulges can show stellar bars
in their central regions. 
The vertical velocity dispersion at a given radius is set to be 0.5
times as large as the radial velocity dispersion at that point,
as is consistent with the observed trend of the Milky Way.

We allocate metallicity to each disk and bulge star 
according to its initial position:
at $r$ = $R$,
where $r$ ($R$) is the projected distance (in units of kpc)
from the center of the disk, the metallicity of the star is given as:
\begin{equation}
{\rm [m/H]}_{\rm r=R} = {\rm [m/H]}_{\rm d, r=0} + {\alpha}_{\rm d} \times {\rm R}. \;
\end{equation}
We adopt the observed  value  of ${\alpha}_{\rm d} \sim -0.04$
(e.g., Andrievsky et al. 2004) for all models,
and the central metallicity ${\rm [m/H]}_{\rm d, r=0}$
is chosen according to the adopted $f_{\rm g}$ and $M_{\rm gal}$.
For the MW model,  ${\rm [m/H]}_{\rm d, r=0}=0.34$, which gives
a reasonable value of [Fe/H]=0 at the solar radius ($R=8.5$ kpc).
The initial central metallicity $Z_0$ (or [Fe/H]$_0$) is chosen according
to (i) the mass metallicity relation of $Z_0 \propto M_{\rm d}^{0.25}$
and (ii) the standard prediction from one-zone chemical evolution
models on the relation between the gas mass fraction ($f_{\rm g}$)
and the metallicity 
of a galaxy (i.e., $Z_0 \propto 
1 + \frac{ f_{\rm g} \ln f_{\rm g} }{1-f_{\rm g}}$).
The initial [$\alpha$/Fe] ratios are set to be the same as the solar values
(e.g., [Mg/Fe]=0) across the stellar and gas disks of a galaxy.
Initial temperature of gas is set to be $10^4$K for all models.

We mainly investigate the MW models  in which $f_{\rm dm}=15.2$,
$c=10$, $R_{\rm vir}=245$ kpc, 
$M_{\rm d}$ = 6.6 $\times$ $10^{10}$ $ \rm
M_{\odot}$,  $R_{\rm d}$ = 17.5 kpc,
$f_{\rm b}=0.15$, $R_{\rm b}=2$ kpc,
and different $f_{\rm g}$ (=0.09, 0.27, and 0.55).
The  MW model with $f_{\rm g}=0.09$ and without tidal interaction
is referred to as `the fiducial model' and the basic
parameters are shown in Table 1.
We also investigate how the present results depend on model
parameters such as  $M_{\rm gal}$ and
$f_{\rm g}$, and presence or absence of tidal interaction.
The model $f_{\rm g}=0.09$ and 0.55 are referred to as `gas-poor' and
`gas-rich' models, respectively.
The model with $M_{\rm d}=6.0 \times 10^9 {\rm M}_{\odot}$,
$f_{\rm g}=0.55$, and $f_{\rm b}=0$ (i.e., no bulge) is referred
to as the LMC (Large Magellanic Cloud) model (M5).
The low-mass  model with $M_{\rm d}=6.0 \times 10^8 {\rm M}_{\odot}$,
$f_{\rm g}=0.55$, and $f_{\rm b}=0$  is referred
to as the dwarf  model (M6). We adopt rather high  baryonic fractions for low-mass disks
(the same as those in the MW models), mainly because we try to understand 
more clearly how the initial total mass can influence IMF evolution by comparing between
the dwarf, LMC, and MW  models with the same initial conditions (other than total masses).

The total numbers of particles used for dark matter, stellar disk,
gas disk, and bulge in an isolated disk are 
700000, 200000, 100000, and 33400, respectively (i.e., $N=1033400$
is used in total).
The softening length of dark matter halo ($\epsilon_{\rm dm}$) for each model
is chosen so that $\epsilon_{\rm dm}$ can be the same as the
mean particle separation at the half-mass radius of the halo.
This method is applied  for 
determining softening length for stellar particles
($\epsilon_{\rm s}$) in the initial disk of each model.
The softening length is assumed to be the same between old stellar,
gaseous, and new stellar
particles in the present study.
The gravitational softening length for dark ($\epsilon_{\rm dm}$)
and baryonic components ($\epsilon_{\rm s}$) 
are 2.1 kpc and 200 pc, respectively, for the MW model.
These values are different in models with different sizes and masses.

\subsection{Tidal interaction}

As shown in BM13, the high-mass end of the 
IMF can become top-heavy during galaxy interaction
and merging.  We accordingly investigate 
how galaxy evolution during galaxy interaction between two spiral galaxies
can be influenced by the  NUIMF. 
One of the two galaxies (`primary galaxy')
is represented by the spiral galaxy model
described above whereas the interacting companion galaxy is represented
by a point-mass particle. 
The mass-ratio of the companion to the primary is a free parameter represented
by $m_2$ and we investigate the models with $m_2=0.3$, 1, and 3. The orbits
of interacting galaxies are assumed to be `prograde interaction' in which
the spin axis of the primary is parallel to the orbital spin axis. 
The initial distance of the two galaxies,  the pericenter distance,
and the orbital eccentricity
are set to be $16R_{\rm d,s}$, $2R_{\rm d,s}$, and 1.0 (i.e., parabolic),
respectively.
A more dramatic change in the time evolution
of gas density and star formation can occur in the adopted prograde
interaction than in other orbital configurations (e.g., retrograde interaction).
Accordingly,  the influences of the NUIMF on interacting galaxies can be more clearly 
seen in the present study (i.e., only the models in which the IMF influences are pronounced
are shown). Thus, it should be stressed that the influences of the NUIMF
on interacting galaxies are not always
seen in the tidal models.

\subsection{Star formation and SN feedback effects}

Since SF can proceed in molecular clouds,
we adopt the following `${\rm H_2}$-dependent' SF recipe 
(B13) using molecular gas fraction
($f_{\rm H_2}$) defined for each gas particle in the present study.
A gas particle {\it can be} converted
into a new star if (i) the local dynamical time scale is shorter
than the sound crossing time scale (mimicking
the Jeans instability) , (ii) the local velocity
field is identified as being consistent with gravitationally collapsing
(i.e., div {\bf v}$<0$),
and (iii) the local density exceeds a threshold density for star formation ($\rho_{\rm th}$).
We mainly investigate the models with $\rho_{\rm th}=1$ cm$^{-3}$
in the present study.

A gas particle can be regarded as a `SF candidate' gas particle
if the above three SF conditions (i)-(iii) are satisfied.
It could be possible to convert some fraction ($\propto f_{\rm H_2}$)
of a SF candidate  gas particle
into a new star at each time step until the mass of the gas particle
becomes very small. However, this SF conversion method can increase dramatically
the total number of stellar particles, which becomes  numerically very costly.
We therefore adopt the following SF conversion method.
A SF candidate $i$-th gas
particle is regarded as having  a SF probability ($P_{\rm sf}$);
\begin{equation}
P_{\rm sf}=1-\exp ( -C_{\rm eff} f_{\rm H_2} 
\Delta t {\rho}^{\alpha_{\rm sf}-1} ),
\end{equation}
where $C_{\rm eff}$ corresponds to a star formation  efficiency (SFE) 
in molecular cores and is set to be 1,
$\Delta t$ is the time step width for the gas particle, 
$\rho$ is the gas density of the particle,
and $\alpha_{\rm sf}$ is
the power-law slope of the  Kennicutt-Schmidt law 
(SFR$\propto \rho_{\rm g}^{\alpha_{\rm sf}}$;  Kennicutt 1998).
A reasonable value of 
$\alpha_{\rm sf}=1.5$ is adopted in the present 
study.

At each time step   random numbers ($R_{\rm sf}$; $0\le R_{\rm sf}  \le 1$)
are generated and compared with $P_{\rm sf}$.
If $R_{\rm sf} < P_{\rm sf}$, then the gas particle can be converted into
a new stellar one.
In this SF recipe, a gas particle with a higher gas density
and thus a shorter SF timescale ($\propto
\rho/\dot{\rho} \propto \rho^{1-\alpha_{\rm sf}}$) 
can be more rapidly converted into a new star owing to the larger
$P_{\rm sf}$. Equally, a gas particle with a higher $f_{\rm H_2}$
can be more rapidly converted into a new star.
We thus consider that the present SF model is a good approximation
for star formation in molecular gas of disk galaxies.

Each SN is assumed to eject the feedback energy ($E_{\rm sn}$)
of $10^{51}$ erg and 90\% and 10\% of $E_{\rm sn}$ are used for the increase
of thermal energy (`thermal feedback')
and random motion (`kinematic feedback'), respectively.
The thermal energy is used for the `adiabatic expansion phase', where each SN can remain
adiabatic for a timescale of $t_{\rm adi}$.
We adopt $t_{\rm adi}=10^5$ yr, which is reasonable for a single SN explosion.
This $t_{\rm adi}$ can be different  
for multiple SN explosions in a small local region owing to complicated
interaction between gaseous ejecta from different SNe.  However, we show the results
of the models with $t_{\rm adi}=10^5$ yr in the present study.
The energy-ratio of thermal to kinematic feedback is consistent with
previous numerical simulations by Thornton et al. (1998) who investigated
the energy conversion processes of SNe in  detail.
The way to distribute $E_{\rm sn}$ of SNe among neighbor gas particles
is the same as described in Bekki et al. (2012).
The radiative cooling processes
are properly included  by using the cooling curve by
Rosen \& Bregman (1995) for  $100 \le T < 10^4$K
and the MAPPING III code
for $T \ge 10^4$K
(Sutherland \& Dopita 1993).

\begin{figure*}
\psfig{file=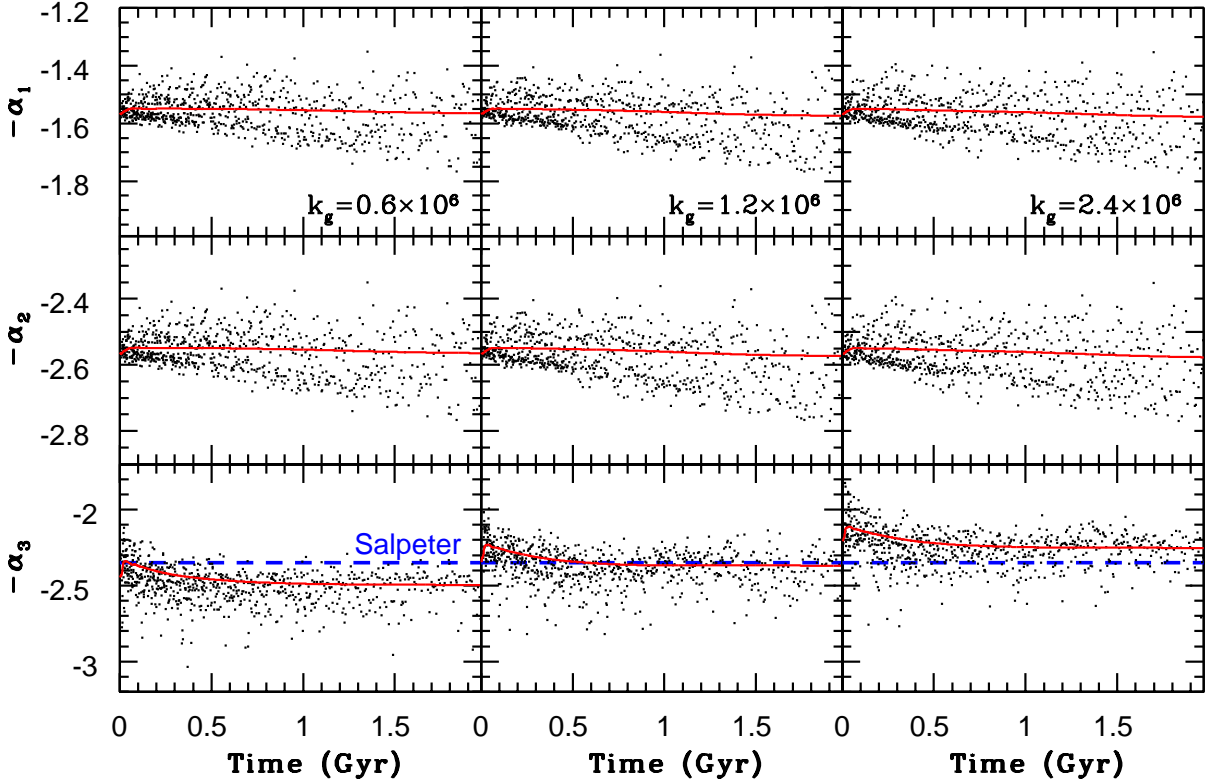,width=16.0cm}
\caption{
The time evolution of $\alpha_1$ (top), $\alpha_2$ (middle),
and $\alpha_3$ (bottom) for the fiducial MW model (M1) with the NUIMF
and three different $k_{\rm g}$: 
$0.6 \times 10^6$ (left),
$1.2 \times 10^6$ (middle),
and $2.4 \times 10^6$ (right).
The red lines indicate the mean values of the IMF slopes at each time step,
and the IMF slopes of new stars formed at each time step are shown by 
small black dots. One every 50 particles is shown so that the figure size
can be dramatically reduced (yet the trend of the IMF evolution
can be clearly seen).
The values of 
$-\alpha_1$, $-\alpha_2$, and $-\alpha_3$
(rather than $\alpha_1$, $\alpha_2$, and $\alpha_3$)  are shown so that the locations
of new stars in the upper part of each frame can indicate more top-heavy IMFs.
For comparison, the Salpeter IMF for $\alpha_3$ is shown by a thick dotted line.
}
\label{Figure. 1}
\end{figure*}

\subsection{Chemical evolution}

Chemical enrichment through star formation and metal ejection from
SNIa, II, and AGB stars is considered to proceed locally and inhomogeneously.
SNe and AGB stars are the production sites of dust, and some metals ejected from
these stars can be also accreted onto dust grains in the ISM of galaxies.
We investigate the time evolution of the 11 chemical elements of H, He, C, N, O, Fe,
Mg, Ca, Si, S, and Ba in order to predict both chemical abundances and 
dust properties
in the present study. 
The mean metallicity $Z$ for each $k$th stellar particle is
represented by $Z_k$. The total mass of each $j$th ($j=1-11$) chemical component
ejected from each $k$th stellar particles at time $t$ is given as
\begin{equation}
\Delta z_{k,j}^{\rm ej}(t)=m_{\rm s, \it k} Y_{k,j}(t-t_k),
\end{equation}
where $m_{\rm s, \it k}$ is the mass of the $k$th stellar particle, $Y_{k,j}(t-t_
k)$
is the mass of each $j$th chemical component ejected from stars per unit mass at
time $t$, and $t_k$ represents the time when the $k$th stellar particle is
born from a gas particle. 
This $Y_{k,j}(t-t_k)$ includes both the metals recycled 
and those newly synthesized, and both metals can be calculated for each stellar
particle  based on the adopted  
yield tables (described below) and the metal abundances of the particle.
$\Delta z_{k,j}^{\rm ej}(t)$ is 
given equally  to neighbor
SPH gas
particles (with the total number of $N_{\rm nei, \it  k}$) 
located around the $k$th stellar particle.  Therefore,
the mass increase of each $j$th chemical component for $i$th gas particle at time
 $t$
($\Delta z_{i,j}^{\rm ej}(t)$) is
given as
\begin{equation}
\Delta z_{i,j}^{\rm ej}(t) = \sum_{k=1}^{N_{\rm nei, \it i}}
m_{\rm s, \it k} Y_{k,j}(t-t_k)/N_{\rm nei, \it k},
\end{equation}
where $N_{\rm nei, \it i}$ is the total number of neighbor stellar particles whose metals
can be incorporated into the $i$th gas particle.

We consider the time delay between the epoch of star formation
and those  of supernova explosions and commencement of AGB phases (i.e.,
non-instantaneous recycling of chemical elements).
Therefore, the mass of each $j$th chemical component ejected from each
$i$th stellar particle is strongly time-dependent.
We also adopt the `prompt SN Ia' model in which
the delay time distribution 
of SNe Ia is consistent with  recent observational results by  
extensive SN Ia surveys (see B13 for the detail of the prompt SN Ia model).
The chemical yields adopted in the present study are the same as those used in
Bekki \& Tsujimoto (2012) except those from AGB stars.
We adopt the nucleosynthesis yields of SNe II and Ia from Tsujimoto et al. (1995; T95)
and AGB stars from van den Hoek \& Groenewegen (1997; VG97)
in order to estimate $Y_{k,j}(t-t_k)$ in the present study.
We mainly investigate [Mg/Fe]$-$[Fe/H] relation of the simulated
galaxies, mainly because this relation
can represent [$\alpha$/Fe]$-$[Fe/H] relations.

\subsection{Dust model}

\subsubsection{Yield}

The present dust model is essentially the same as that adopted in the previous
multi-zone model by Dwek (1998, D98), which reproduced reasonably well the observed
chemical and dust properties of the Galaxy in a self-consistent manner.
The dust model consists of the following four components: (i) production in stellar winds
of SNe Ia and SNe II and AGB stars,  (ii) accretion of metals of ISM on dust grains,
(iii) destruction of dust by energetic SN explosions, and (iv) PAH formation.
The present model is somewhat idealized in that it does not include coagulation
of small dust grains and time evolution of dust sizes.

The total mass of $j$th component ($j$=C, O, Mg, Si, S, Ca, and Fe)
of dust from $k$th type of stars ($k$ = I, II, and AGB for SNe Ia, SNe II, and
AGB stars, respectively) is described as follows;
\begin{equation}
m_{\rm dust, \it j}^k= \delta_{\rm c, \it j}^k F_{\rm ej}(m_{\rm ej, \it j}^k),
\end{equation}
where $\delta_{\rm c, \it, j}^k$ is the condensation efficiency (i.e., the mass
fraction of metals that are locked up in dust grains) for each $j$th 
chemical component
from $k$th stellar type,
$F_{\rm ej}$ is the function that determines the total mass of metals that can be
used
for dust formation,
and $m_{\rm ej, j}^k$ is the mass of $j$th component
ejected from $k$th stellar type.
The total mass of stellar ejecta is estimated by using stellar yield tables 
by T95
and VG97.
We adopt the exactly same 
$\delta_{\rm c, \it j}^k$ and $F_{\rm ej}(m_{\rm ej, \it j}^k)$ 
as those used in B13.

\subsubsection{Accretion}

Dust grains can grow by accretion of metals of ISM onto preexisting cores and this
accretion process is included in previous models (D98). Following D98, we consider
that the key parameter in dust accretion is the dust accretion timescale ($\tau_{\rm acc}$).
In the present study, this parameter can vary between different gas particles
and is thus represented by $\tau_{\rm acc, \it i}$ for $i$th gas particle.
The mass of $j$th component
($j$=C, O, Mg, Si, S, Ca, and Fe) of dust for $i$th gas particle
at time $t$ ($d_{i,j}(t)$) can increase owing  to dust accretion processes.
The mass increase
is described as
\begin{equation}
\Delta d_{i,j}^{\rm acc}(t)=\Delta t_i (1-f_{\rm dust,\it i, j})
d_{i,j}(t) /\tau_{\rm acc, \it i},
\end{equation}
where $\Delta t_i$ is the individual time step width for the $i$th gas particle
and $f_{\rm dust, \it i, j}$ is the fraction of the $j$th chemical element that
is locked up in the dust. Owing to this dust growth, the mass of $j$th chemical
component that is {\it not} locked up in the dust ($z_{i,j}(t)$)
can decrease, which is simply given as
\begin{equation}
\Delta z_{i,j}^{\rm acc}(t)=- \Delta t_i (1-f_{\rm dust,\it i, j})
d_{i,j}(t) /\tau_{\rm acc, \it i}
\end{equation}
As is clear in these equations, the total  mass of $j$th component in $i$th gas
particle ($m_{i,j}(t)$) is $z_{i,j}(t)+d_{i,j}(t)$.
For all models, $\tau_{\rm acc}$ is set to be 0.25 Gyr in the present study.

\begin{figure*}
\psfig{file=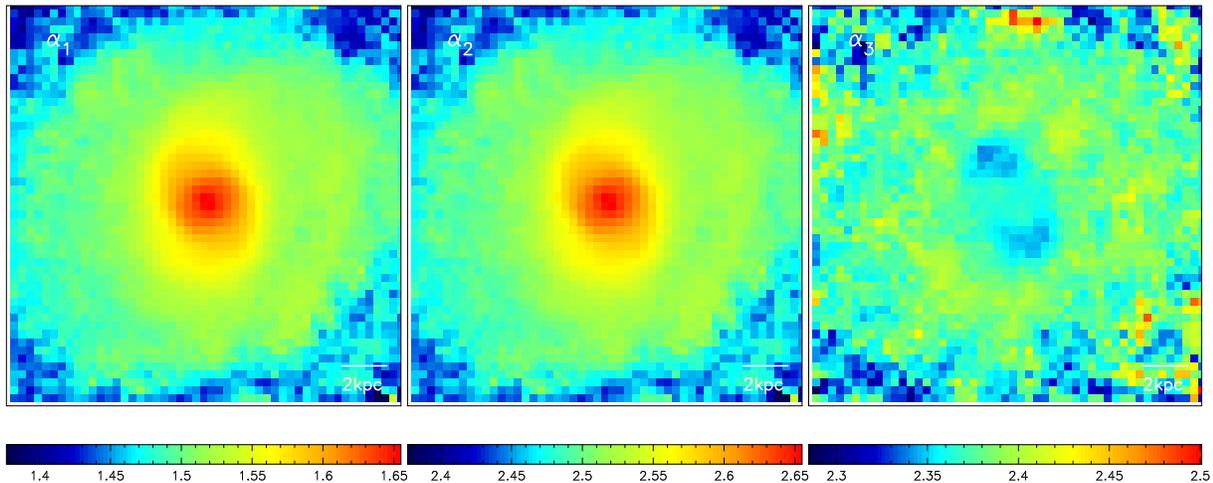,width=16.0cm}
\caption{
The 2D distributions of $\alpha_1$ (left), $\alpha_2$ (middle),
and $\alpha_3$ (right) of the isolated disk galaxy
projected onto the $x$-$y$ plane
at $T=2$ Gyr in the fiducial model M1. 
The simulated region is divided $100 \times 100$ cells and the mean
IMF slopes are estimated for each cell. 
}
\label{Figure. 2}
\end{figure*}

\begin{figure*}
\psfig{file=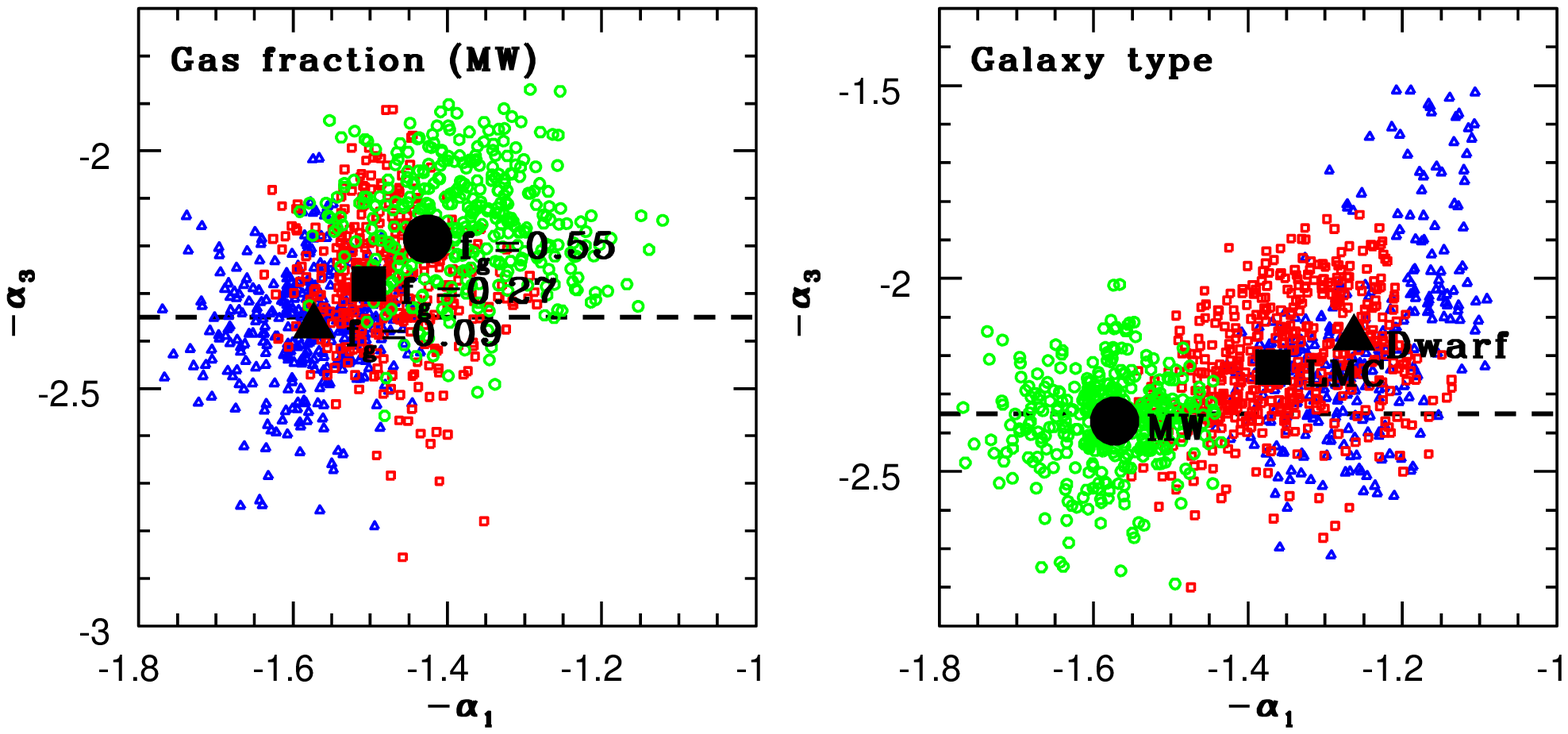,width=16.0cm}
\caption{
The locations of star-forming regions on
the $\alpha_1-\alpha_3$ plane for the three MW models (left)
with $f_{\rm g}=0.09$ (M1, blue triangle),
$f_{\rm g}=0.27$ (M2, red square),
and MW with $f_{\rm g}=0.55$ (M3, green circle)
and for the three different galaxy models (right),
dwarf ((M6, blue triangle),
LMC (M5, red square),
and  MW  (M1, green circle).
For comparison, the Salpeter IMF slope ($\alpha_3=2.35$) is shown by a dashed
line in each panel.
}
\label{Figure. 3}
\end{figure*}

\subsubsection{Destruction}
Dust grains can be destroyed though supernova blast waves
in the ISM of galaxies (e.g., McKee 1989)
and the destruction process is parameterized by the destruction time scale
($\tau_{\rm dest}$) in previous one-zone models (e.g., Lisenfeld \& Ferrara 1998;
Hirashita 1999).  Following the previous models,
the decrease  of the mass of $j$th component
of dust for $i$th gas particle
at time $t$ due to dust destruction process
is as follows
\begin{equation}\Delta d_{i,j}^{\rm dest}(t)= - \Delta t_i
d_{i,j}(t) /\tau_{\rm dest, \it i},
\end{equation}
where $\tau_{\rm dest, \it i}$ is the dust destruction timescale for $i$th 
particle.
The dust destroyed by supernova explosions can be returned back to the ISM,
and therefore the  mass
of $j$th chemical
component that is not locked up in the dust
increases  as follows:
\begin{equation}
\Delta z_{i,j}^{\rm dest}(t)= \Delta t_i
d_{i,j}(t) /\tau_{\rm dest, \it i}
\end{equation}

Thus the equation for the time evolution of $j$th component of metals
for $i$th gas particle  are given as
\begin{equation}
z_{i,j}(t+\Delta t_i)=z_{i,j}(t)+\Delta z_{i,j}^{\rm ej}(t)+\Delta z_{i,j}^{\rm acc}(t)
+\Delta z_{i,j}^{\rm dest}(t)
\end{equation}
Likewise, the equation for dust evolution is given as
\begin{equation}
d_{i,j}(t+\Delta t_i)=d_{i,j}(t)+\Delta d_{i,j}^{\rm acc}(t)
+\Delta d_{i,j}^{\rm dest}(t)
\end{equation}
 Dust is locked up in stars as metals are done so, when gas particles are converted into
new stars. This means that star formation process itself has an effect
of destroying dust in the present study.
As shown in B13, models with $\tau_{\rm acc}/\tau_{\rm dest}=0.5$ can explain
the dust-to-gas ratio ($D$) in luminous disk galaxies.
Therefore $\tau_{\rm dest}$ is set to be 0.5 Gyr in the present study.

\subsection{${\rm H_2}$ formation and dissociation}

The model for ${\rm H_2}$ formation and dissociation in the present study
is exactly the same as those used in B13: ${\rm H_2}$ formation 
on dust grains and ${\rm H}_2$ dissociation by FUV radiation
are both self-consistently included in chemodynamical simulations. 
The temperature ($T_{\rm g}$),  
hydrogen density ($\rho_{\rm H}$),  dust-to-gas ratio ($D$)
of a gas particle and the strength of the 
FUV radiation field ($\chi$) around the gas particle
are calculated at each time step so that the fraction of molecular
hydrogen ($f_{\rm H_2}$) for the gas particle can be derived based on
the ${\rm H_2}$ formation/destruction equilibrium conditions.
Thus the ${\rm H_2}$ fraction for $i$-th gas  particle ($f_{\rm H_2, \it i}$)
is given as;
\begin{equation}
f_{\rm H_2, \it i}=F(T_{\rm g, \it i}, \rho_{\rm H, \it i}, D_i,  \chi_i),
\end{equation}
where $F$ means a function for $f_{\rm H_2, \it i}$ determination,
and the detail of the derivation process of $f_{\rm H_2}$ are given
in B13. 
Thus each gas particle has $f_{\rm H_2}$, metallicity ([Fe/H]),
and gas density, all of which are used for estimating the IMF slopes 
for the particle (when it is converted into a new star).

\begin{figure*}
\psfig{file=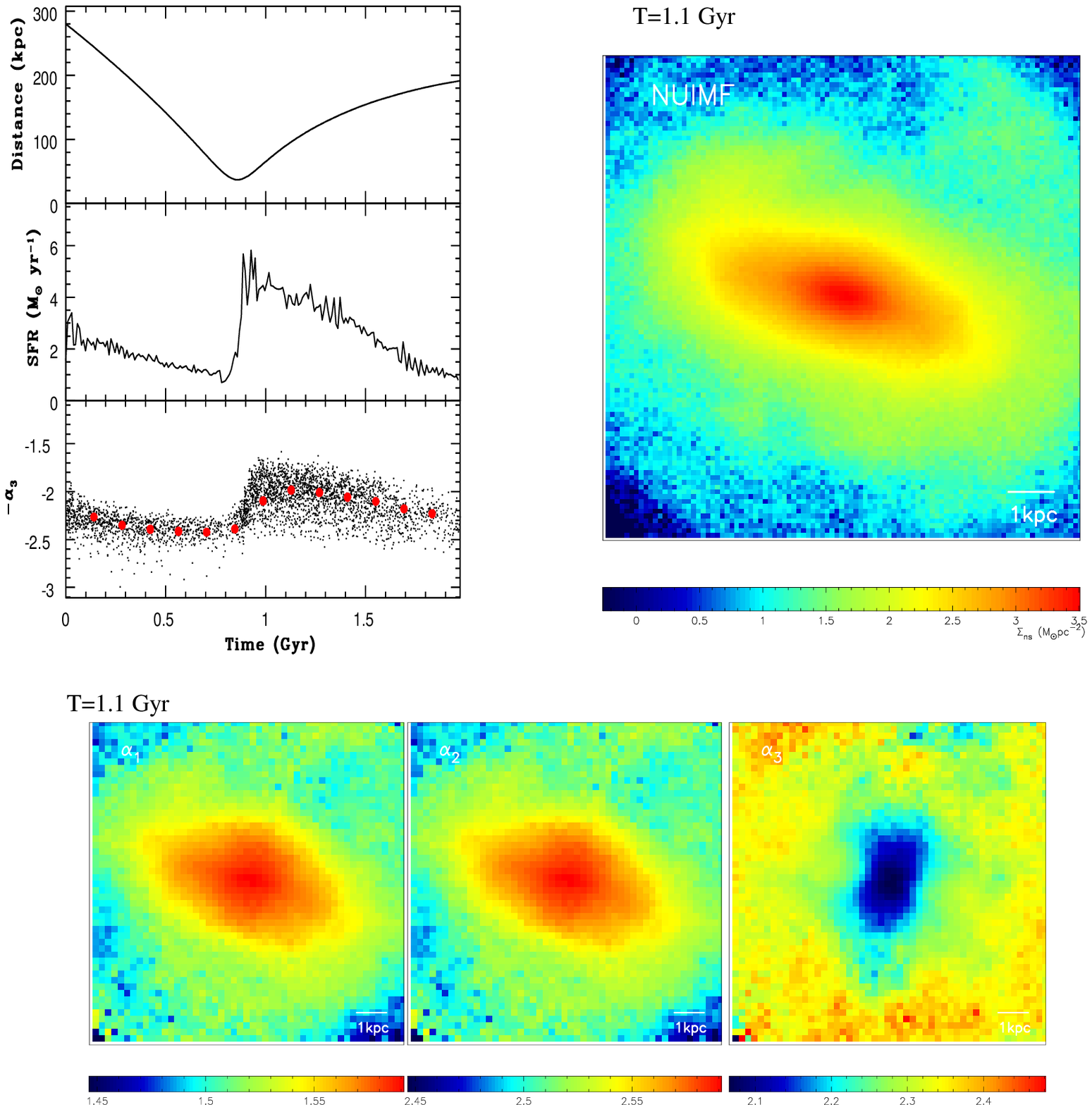,width=14.0cm}
\caption{
A collection of three figures describing the evolution of
the IMF and SFR (upper left), the 2D distribution of projected mass densities
of new stars ($\Sigma_{\rm ns}$)
at the strong starburst phase ($T=1.1$ Gyr) triggered
by tidal interaction (upper right), and the 2D distributions of the IMF slopes
at $T=1.1$ Gyr (bottom)
in the tidal interaction model M4.
In the upper left panel, the time evolution of the distance of two galaxies
(top), SFR (middle), and $-\alpha_3$ (bottom)
are shown. In this panel,  $-\alpha_3$ for one every 20 new stellar particles
formed at each time step is shown by small black dots and
the mean values of $\alpha_3$ at selected time steps are shown to describe
the evolution of the mean $\alpha_3$ in the interacting galaxy more clearly.
}
\label{Figure. 4}
\end{figure*}

\subsection{The varying  Kroupa IMF model}

We adopt the NUIMF model proposed by M12
in which (i) the basic multi-power-law form is the same as
the Kroupa IMF (Kroupa 2000)  and 
(ii)  the IMF slopes ($\alpha_{i}$, {\it i}=1, 2, 3) depend on
the densities and metallicities of gas clouds from which new stars
form. 
Although we choose the M12 IMF  to
discuss how galaxy evolution is influenced by the NUIMF in the
present study, this choice would not be 
regarded as the most reasonable
and realistic. Recently Narayanan \& Dav\'e (2012) adopted a NUIMF
model in which the characteristic mass of stars is proportional
to ${\rm (SFR)}^{0.3}$, where SFR is the star formation rate of a local region
in a galaxy.
Their IMF model  is therefore significantly different 
from the M12's  IMF in the sense that the IMF slope does not change
in their model (see their Fig.2).
We do not discuss which of the two IMFs is better and realistic in terms
of reproducing the observed properties of galaxies in the present study.

The Kroupa IMF has the following form (M12);
\begin{equation}
dN/dm=\xi(m)=ka_i \times m^{-\alpha_i},
\end{equation}
where $N$ is the number of stars forming in the mass interval [$m$, $m+dm$],
$\alpha_i$ ($i=$1, 2, 3) is the power-law
slope of the IMF, 
$k$ is a normalization constant,  
and $a_i$ is a constant that should be chosen for a given
set of $a_i$ to warrant
continuity at the edges of the power-law segments.
For the canonical IMF, $\alpha_1$, $\alpha_2$, and $\alpha_3$ are 1.3,
2.3, and 2.3, respectively, for
$0.08 \le m/{\rm M}_{\odot} < 0.5$,
$0.5 \le m/{\rm M}_{\odot} < 1$,
and $1 \le m/{\rm M}_{\odot} \le m_{\rm max}$, where
$m_{\rm max}$ is the stellar upper-mass limit and
set to be $100 {\rm M}_{\odot}$ in the present study.

The low-mass end of the Kroupa IMF for each
$i$-th new stellar particle ($\alpha_{1,i}$) depends solely on [Fe/H]
as follows
\begin{equation}
\alpha_{1, i}=1.3+0.5\times {\rm [Fe/H]}_i,
\end{equation}
where [Fe/H]$_i$ is the iron abundance of the particle.
The value of $\alpha_2$ for
$i$-th new stellar particle ($\alpha_{2,i}$) is also determined
solely by [Fe/H]; 
\begin{equation}
\alpha_{2, i}=2.3+0.5\times {\rm [Fe/H]}_i.
\end{equation}
The high-mass end of the Kroupa IMF for each
$i$-th new stellar particle ($\alpha_{3,i}$) is described as follows;
\begin{equation}
\alpha_{3,i} = 0.0572 \times {\rm [Fe/H]}_i -0.4072 \times \log_{10}(
\frac{ \rho_{\rm cl, \it i} }{ 10^6 M_{\odot} {\rm pc}^{-3} }) +1.9283,
\end{equation}
where $\rho_{\rm cl, \it i}$ is the density of a rather high-density  gaseous
core where star formation can occur.
This equation holds for $x_{\rm th} \ge -0.87$, where
$x_{\rm th}=-0.1405{\rm [Fe/H]}+\log_{10}(
\frac{ \rho_{\rm cl} }{ 10^6 M_{\odot} {\rm pc}^{-3} })$,
and $\alpha_{3,i}=2.3$ for $x_{\rm th} <-0.87$ (M12).  We slightly modify
the M12's IMF model such that
the threshold $x_{\rm th}$ in M12 is not introduced in the present study.
This is 
mainly because the $\alpha$-dependence at $x < x_{\rm th}$ is not
so clear (not so flat as M12 showed) owing to a small number of data points.

\begin{figure*}
\psfig{file=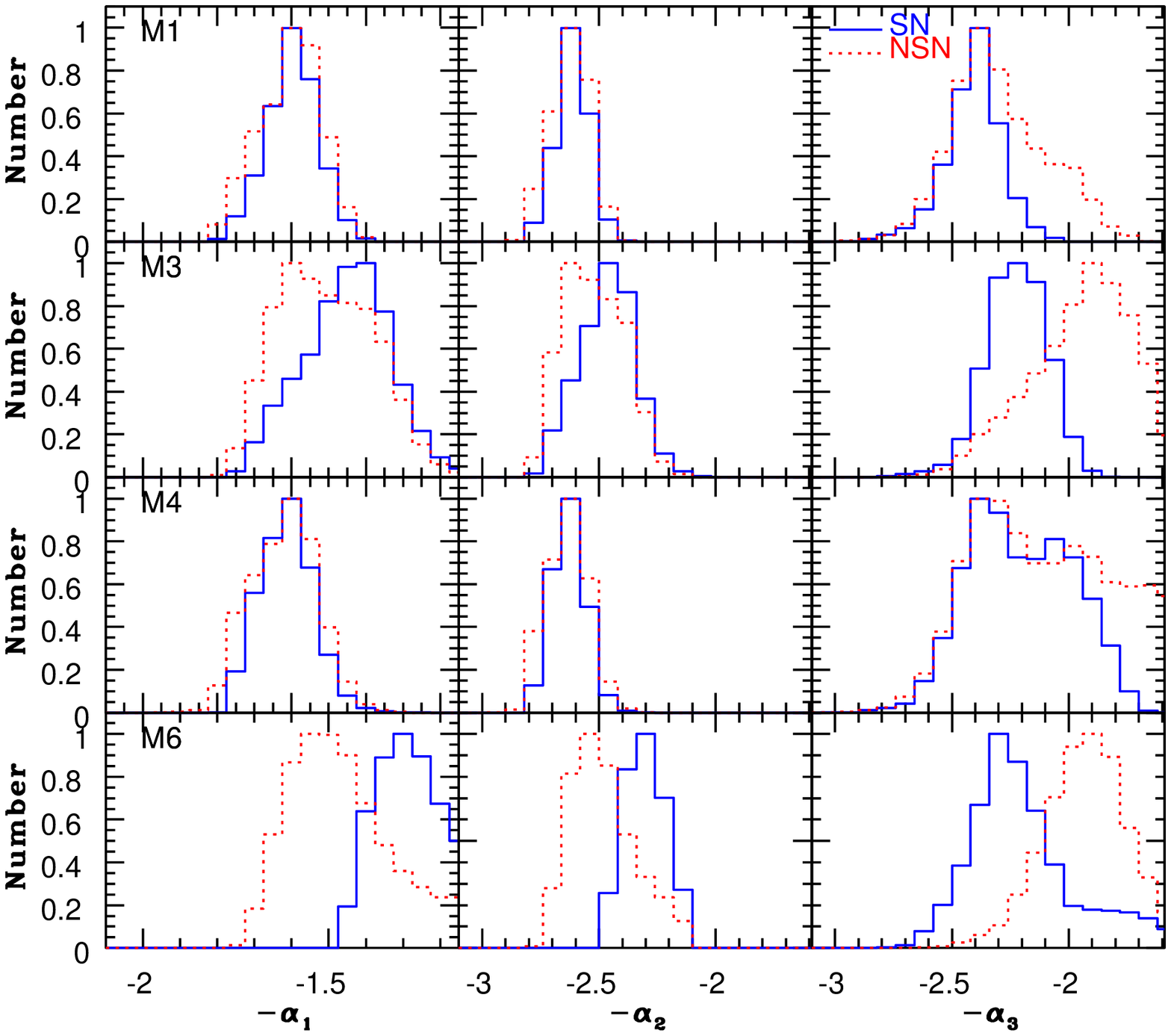,width=16.0cm}
\caption{
The histograms of the three IMF slopes for the four different models,
M1 (top), M3 (the second from top), and M4 (the second from bottom),
and M6 (bottom)
with (blue solid) or without (red dotted)
SN feedback effects (`SN' and `NSN', respectively).
`Number' in the $y$-axis for these plots
is `normalized number' in the sense that the number of new stars at each
IMF bin for each IMF slope (e.g., $\alpha_1$)
is divided by the maximum number among all bins for the IMF slope. 
}
\label{Figure. 5}
\end{figure*}

We can directly estimate  $\alpha_{1,i}$ and $\alpha_{2,i}$ by using 
[Fe/H] and the above equations (14) and (15).
We estimate  $\rho_{\rm cl}$ from local gas density $\rho_{\rm g}$
and $f_{\rm H_2}$
of Jeans-unstable gas particles
in  equation (16).  
This is because the present simulation can not resolve
the rather high-density cores of star-forming or cluster-forming
molecular gas clouds.
The local value of $\rho_{\rm cl}$ at each Jeans-unstable gas is
estimated by multiplying the mean ${\rm H_2}$ density of a cloud,
$\rho_{\rm H_2}$ ($ \approx f_{\rm H_2} \times  \rho_{\rm g}$), 
by a constant $k_{\rm g}$:
\begin{equation}
\rho_{\rm cl}=k_{\rm g}\rho_{\rm H_2}.
\end{equation}
The derivation process of $k_{\rm g}$ is as follows.
The Jeans-unstable gas is assumed to be star-forming giant molecular
clouds and thus have the following  size-mass scaling relation
derived from the observed mass-density relation by Larson (1981):
\begin{equation}
R_{\rm gmc}=40 \times (\frac{ M_{\rm gmc} }{ 5\times 10^5 M_{\odot} })^{0.53} {\rm pc}.
\end{equation}
Since the equation  (16) is based largely on the observed properties of the Galactic
GC
and nearby star clusters,
a reasonable  $k_{\rm g}$ can be the typical density ratio of GCs to GC-host GMCs.
We here  consider that (i) $\rho_{\rm cl}$ should correspond to a typical mean mass
density for GCs,
(ii) typical GC mass ($M_{\rm gc}$) and size ($R_{\rm gc}$) are $2 \times 10^5 M_{\odot}$
and 3pc, respectively (Binney \& Tremaine 2007),
(iii) original GCs just after  their formation from GMCs
should be $\sim 10$ times more massive than
the present ones  (e.g., Decressin et al. 2010; 
Marks \& Kroupa 2010; Bekki 2011),
and (iv) a star formation efficiency of GC-host
GMCs is $\sim 0.1$.
For $M_{\rm gmc}=2\times 10^7 M_{\odot}$ and $R_{\rm gmc}=283$ pc
in the typical GC-host GMC,
a reasonable $k_{\rm g}$ ($=(R_{\rm gmc}/R_{\rm gc})^{3}$)
is estimated to be  $8.4 \times 10^5$ in BM13.

In the present paper, we adopt  $k_{\rm g}=1.2 \times 10^6$
rather than  the above $k_{\rm g}=8.4 \times 10^5$ in BM13.
We need to adopt $k_{\rm g}$ different from the one adopted in BM13,
firstly because
BM13 did not include
$\rm H_2$ formation model at all (thus less accurately estimate $k_{\rm g}$),
and secondly because the predicted slope for the high-mass end of the IMF 
in the MW model 
could not be so consistent with the observed one of the Galaxy ($\alpha_3 \sim 2.3$) for
$k_{\rm g}=8.4 \times 10^5$. 
We run a number of models with different $k_{\rm g}$ and confirm that
if we   adopt
$k_{\rm g}=1.2 \times 10^6$, then not only the mean $\alpha_3$ of the present
MW model with $f_{\rm g}=0.09$ but also the $\rm H_2$ fraction in the gas disk
can be consistent with the observed ones of the MW.
We therefore adopt
$k_{\rm g}=1.2 \times 10^6$ for all models in the present study.
We discuss these points later in 3.1 by using the results of the models
with different $k_{\rm g}$.

In the present study, $k_{\rm g}$ is assumed to take  the same value for different
galaxies and different environments. This assumption appears to be oversimplified,
given that previous theoretical studies showed the dependences of  physical properties
of atomic and molecular   clouds on metallicity, far-ultraviolet radiation field, and
the ionization rate in the Galaxy (e.g., Wolfire et al, 2003; Krumholz et al. 2009).
In this situation, the best thing that the present study can do is to clearly show  how
the present results could possibly depend on $k_{\rm g}$. This point is briefly discussed
later in \S 3.1. Our future more sophisticated simulations will need to include
the dependences of central ${\rm H_2}$ densities of gas clouds on time-dependent local ionization
rates that are not explicitly included in the present simulations.

Almost all models in the present study
do not show large $\alpha$ ($> 3.1$; very steep IMF) that
is needed to explain the observed  $H_{\alpha}/$FUV flux ratios
of low surface brightness galaxies (LSBs)  in M09.
This  is mainly because star formation can occur only in higher density gaseous regions
where $\alpha$ can be smaller in the present models.
Although the original IMF model by M12 is theoretically derived from observations
on physical properties of GCs and star clusters,
we assume that the IMF model
applies for all new  stars in each bin (i.e., not just for star clusters).

\begin{figure*}
\psfig{file=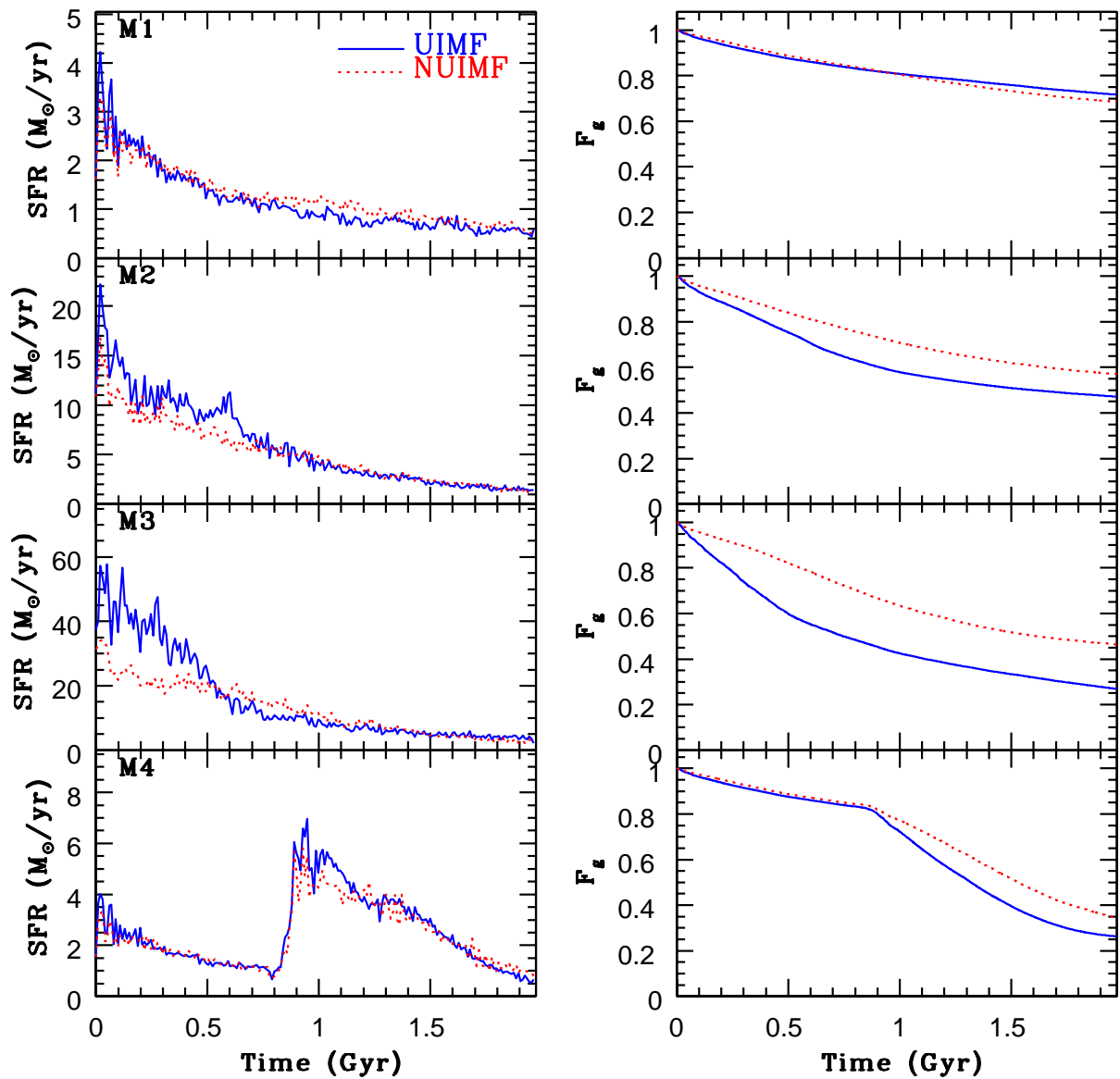,width=16.0cm}
\caption{
The time evolution of SFR (left) and normalized gas mass  $F_{\rm g}$ (right)
for M1 (top), M2 (the second from top),
M3 (the second from bottom), and M4 (bottom) for the UIMF (blue solid)
and the NUIMF (red dotted).
The normalized gas mass at each time step ($F_{\rm g}(t)$)
is defined as $M_{\rm g}(t)/M_{\rm g}(0)$,
where $M_{\rm g}(0)$ is the initial total gas mass (including metals and dust)
of a galaxy. The time evolution of $F_{\rm g}$ can indicate the normalized gas consumption
rate of a galaxy ($R_{\rm gas}$, defined in the main text) in each model.
}
\label{Figure. 6}
\end{figure*}

\subsection{Main points of analysis}

We mainly focus on detailed comparison between models with the UIMF and the NUIMF
in order to understand how star formation histories (SFHs) and
chemical and dynamical evolution of galaxies can be influenced by
a time-varying IMF.
Table 2 summarizes parameter values for
the 15 representative models with different model  parameters
(e.g., $M_{\rm dm}$, $f_{\rm g}$,
and $f_{\rm b}$)
for which the simulation results are described in detail.
The MW models with different $f_{\rm g}$ (M1, M2, and M3),
with the UIMF/NUIMF, and with/without SN feedback effects (`SN' and `NSN',
respectively)
are the most  extensively investigated
so that the physical roles of the NUIMF in galaxy evolution can be clearly
elucidated.

The tidal interaction models (M4, M14, and M15)
are investigated so that we can better understand how the adopted NUIMF
influences  galaxy evolution during strong starbursts triggered by tidal
interaction. 
In order to understand how the IMF roles in galaxy evolution
depend on initial galaxy masses,
we investigate
the LMC models with $M_{\rm dm}=10^{11} {\rm M}_{\odot}$ (M5 and M12)
and
the dwarf models with $M_{\rm dm}=10^{10} {\rm M}_{\odot}$ (M6 and M13).
We also try to find possibly different roles of the NUIMF in
disk galaxies with different bulge masses and mean surface mass densities
(M7, M8, M9, M10, and M11).
In the following,  $T$ in a simulation  represents the time that has elapsed since
the simulation started.

\begin{figure*}
\psfig{file=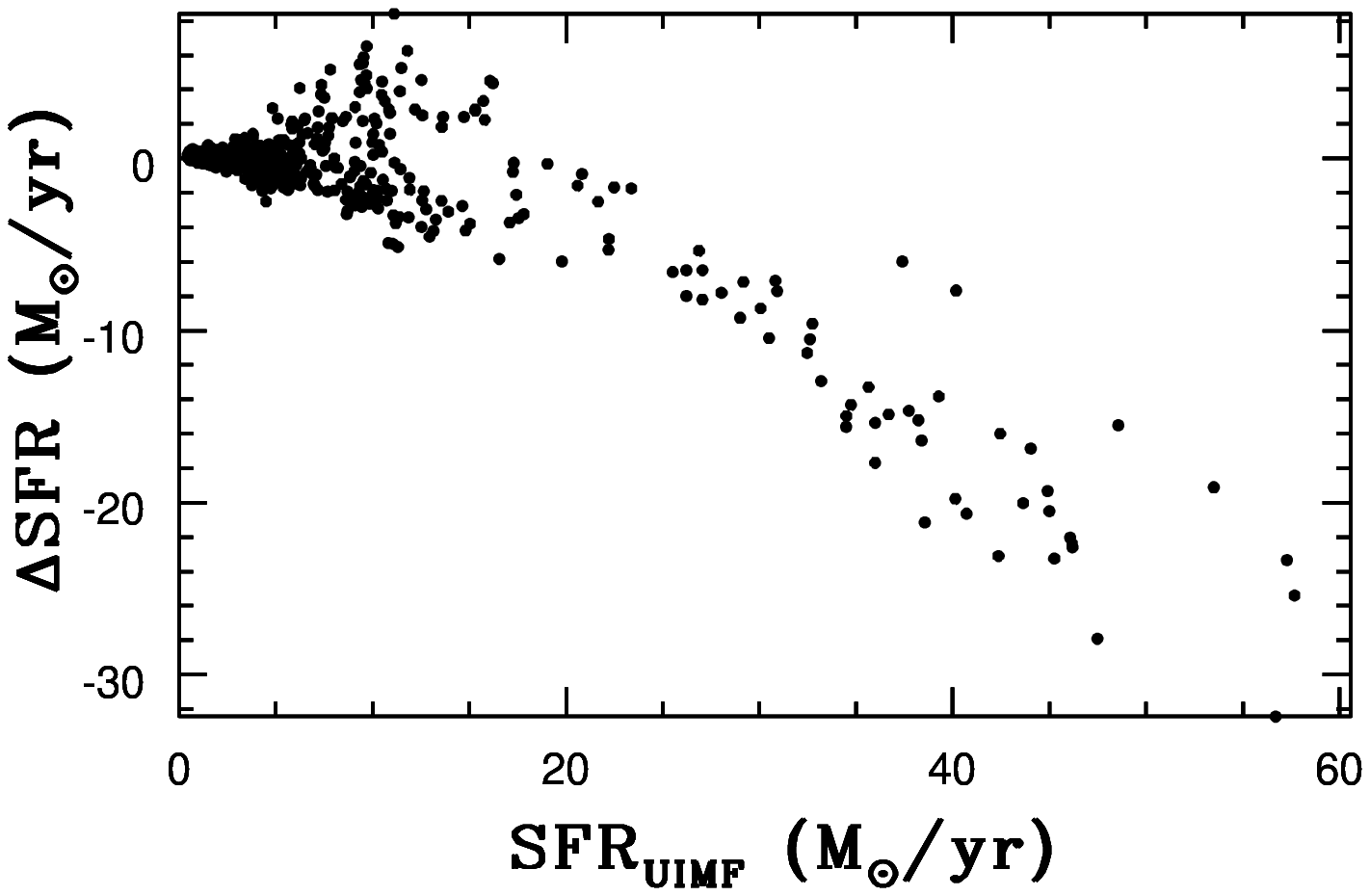,width=8.0cm}
\caption{
The correlation between SFRs in the models with the UIMF (SFR$)_{\rm UIMF}$
and $\Delta {\rm SFR}$ ($={\rm SFR}_{\rm NUIMF}-{\rm SFR}_{\rm UIMF}$)
for the four models shown in Fig. 6.
}
\label{Figure. 7}
\end{figure*}

\section{The predictive power of simulations}

Since there are no previous simulations 
which investigated both the IMF evolution
and its influences on galaxy evolution in a self-consistent manner,
we consider that it is important for this paper to
describe what IMF properties the new chemodynamical simulations
can predict for different galaxies.
Thus we briefly  describe the time evolution of the IMF slopes,
the 2D distributions of the slopes within galaxies, and the physical
connection between top-heavy IMFs and galaxy interaction and merging
before we discuss the differences in galaxy evolution between
the UIMF and NUIMF models in \S 4.

\subsection{IMF evolution}

Fig. 1 shows (i) the formation time and the IMF slopes for each new star
and (ii)  the time evolution of the mean IMF slopes in the 
fiducial  model with $k_{\rm g}=1.2 \times 10^6$
(M1 with the NUIMF model and SN feedback effects).
Different new stars form in gas clouds with different gas densities
and [Fe/H] so that the stars can have different IMF slopes. As a result of this,
the IMF slopes show some dispersion for a given time.
The IMF slope
$\alpha_1$ ($\alpha_2$), which is determined solely by [Fe/H],
ranges from 1.3 (2.3) to 1.8 (2.8) for the last 2 Gyr
disk galaxy evolution, and there is no remarkable evolution of $\alpha_1$
owing to a small degree of chemical enrichment in this gas-poor disk.
The high-mass end of the IMF ($\alpha_3$) does not change during the isolated
disk evolution, though the dispersion is larger than $\alpha_1$ and $\alpha_2$
($1.9 \le \alpha_3 \le 3$). The mean $\alpha_3$ can stay around $2.3 - 2.4$,
which is similar to the canonical IMF. 
The derived no/little evolution of the IMF slopes in the isolated gas-poor
disk model means that the present simulations enable us to investigate
how external perturbation (e.g., galaxy interaction/merging) and 
high gas fraction can influence the IMF evolution (as discussed later).

Fig. 1 also shows the time evolution of the three IMF slopes  in
the models with different $k_{\rm g}$.
Owing to the adopted dependence of $\alpha_3$ on gas density, the time evolution
of $\alpha_3$ depends more strongly on $k_{\rm g}$
in comparison with $\alpha_1$ and $\alpha_2$  and thus can be used to
determine which $k_{\rm g}$ is reasonable and realistic in the present study
in which $k_{\rm g}$ is fixed  in all models. 
Clearly, the model
with $k_{\rm g}=1.2 \times 10^6$ is better at reproducing the Salpeter slope observed in
the solar neighborhood than the  other two models.
 Furthermore,
the evolution of $\alpha_3$ is stable during $\sim 2$ Gyr evolution of the disk.
These results confirm that the present NUIMF model
with $k_{\rm g}=1.2 \times 10^6$  can be used in investigating
chemical and dynamical influences of the NUIMF on galaxies.

\subsection{2D distributions of the IMF slopes}

Fig. 2 shows the 2D distributions of the IMF slopes projected on
the $x$-$y$ plane after 2 Gyr evolution of the disk in the fiducial  model.
The star-forming regions of the disk galaxy
is divided into $100 \times 100$ cells  and the mean IMF slopes 
for each cell  are 
estimated from new stars within each cell
in this figure. The $\alpha_1$ and $\alpha_2$
distributions clearly show radial gradients in the sense that the inner
regions have steeper IMFs. This is simply a reflection of the initial
negative metallicity gradient of the disk, because the radial metallicity
gradient does not evolve so much in this model. 
The $\alpha_3$ distribution, on the other hand, does not show a clear
radial gradient, which reflects the fact that $\alpha_3$ 
depends both on local ${\rm H_2}$ densities and [Fe/H]. 
The high-mass end of the IMF is only
slightly shallower (i.e., slightly more top-heavy) in the central
barred region, which is  due largely to the higher gas density 
of ${\rm H_2}$ in
the barred region. An intriguing feature in the $\alpha_3$ distribution
is a slightly shallower IMF slope in both  edges of the barred region.
Although the origin of 
this feature could be related to the formation of high-density
${\rm H_2}$ regions in the edges of stellar bars, it is beyond the
scope of this paper. We need to investigate this feature in a large number
of simulated barred galaxies in our future studies in order to understand
the origin of this clearly.

\begin{figure*}
\psfig{file=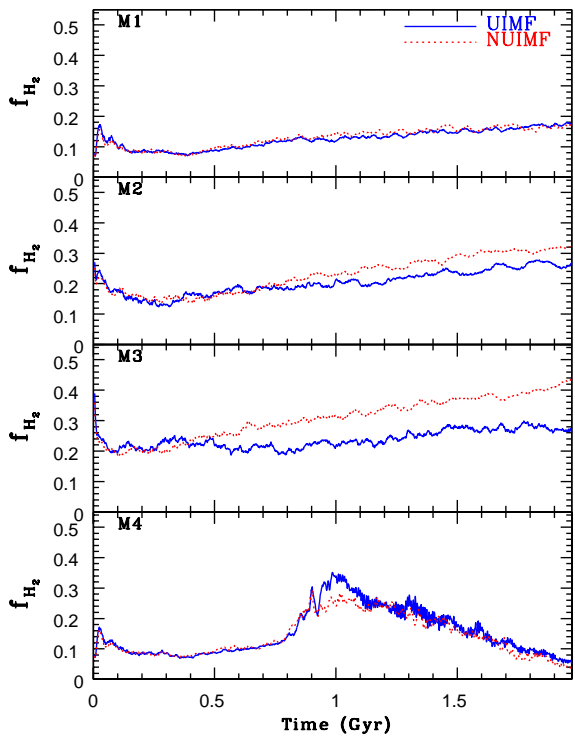,width=8.0cm}
\caption{
The same as Fig. 6 but for $f_{\rm H_2}$.
}
\label{Figure. 8}
\end{figure*}

\subsection{Different IMFs in different galaxies}

Fig. 3 shows the locations of individual star-forming regions 
on the $\alpha_1 - \alpha_3$ plane for the MW models with different $f_{\rm g}$
and for the models with different galaxy types (i.e., dwarf, LMC, and MW).
In this figure,  it is the decreasing [Fe/H] that moves points to the right
in the $x$-axis (see equation (14)).  
On the other hand, it is the gas density that moves points to the top in the $y$-axis
(equation (16); $\alpha_3$ depends rather weakly on [Fe/H]). 
Since $\alpha_2=\alpha_1+1$ in the present study,
the distributions of star-forming regions  on the $\alpha_2 - \alpha_3$ plane are
essentially similar to those on  the  $\alpha_1 - \alpha_3$ plane.

Clearly, more gas-rich MW models show smaller  $\alpha_1$ 
owing to the lower initial [Fe/H], as expected from the adopted NUIMF model.
More strongly self-gravitating gas disks  can cause the formation
of ${\rm H_2}$ clouds with higher densities so that
$\alpha_3$ can become smaller (i.e., more top-heavy) 
in more gas-rich MW models. This result implies that the high-mass end
of the IMF can be more top-heavy in the high redshift universe where
disk galaxies are likely to be more gas-rich.
The gas-rich LMC and dwarf models can  show smaller $\alpha_3$ owing
to the formation of high-density ${\rm H_2}$ gas clouds,
and these low-mass galaxies  can also show smaller $\alpha_1$.
These results imply that (i) low-mass disk galaxies can have top-heavy IMFs
for the entire stellar mass ranges and thus (ii) the top-heavy IMFs could dramatically
influence the early evolution of these galaxies.
We will discuss more extensively
how the IMF slopes depend on physical properties of
galaxies in our forthcoming papers.

\subsection{IMFs in interacting galaxies with starbursts}

Although BM13 demonstrated, for the first time, that
the high-mass end of the IMF can become top-heavy in starbursts
triggered by galaxy interaction and merging,
their simulations are not so fully self-consistent in the sense
that (i) ${\rm H_2}$ evolution is not included and (ii) IMF-dependent
physical effects (e.g., SN feedback) are not properly implemented.
The present model, which is an improved version of the model by BM13,
enables us to discuss better whether the IMF can become top-heavy
during starbursts in interacting and merging galaxies.
Fig. 4 is a collection of plots on the evolution of the SFR and the IMF
slope $\alpha_3$ driven by galaxy interaction and the 2D distributions
of new stars and the three IMF slopes for the tidal interaction model M4.
Clearly, the high-mass end of the IMF can become more top-heavy during
the strong starburst triggered by tidal interaction (i.e., after the pericenter
passage). Because of rapid gas consumption by star formation during
tidal interaction,  the formation of high-density ${\rm H_2}$ gas clouds
can be suppressed after tidal interaction so that
$\alpha_3$ gradually becomes larger (i.e., less top-heavy).
These are therefore consistent with our early results in BM13.

Fig. 4 also shows strong radial gradients of the three IMF slopes
in the central 5.3 kpc at
$T=1.1$ Gyr in the tidal 
interaction model. The IMF slopes $\alpha_1$ and $\alpha_2$
show strong negative gradients, which means that the IMF is more bottom-heavy
in the central few kpc. The IMF slope $\alpha_3$, on the other hand,
shows a positive radial gradient in the sense that the IMF is more top-heavy
in the inner region. The strong radial gradient of $\alpha_3$ 
combined with the weaker $\alpha_3$ gradient
in the isolated MW model (M1)  means that
galaxy interaction can steepen the IMF gradient. Furthermore,
the central region with more top-heavy IMF in $\alpha_3$ forms 
a bar-like structure with the major axis not aligned with the major
axis of the bar shown in the 2D surface density map of new stars
($\Sigma_{\rm ns}$).

This reflects the fact that a secondary bar
can form from gas (transferred into the central region via gaseous dissipation)
within the original bar (i.e., a `nested'  bar formation or
`bar within bar') and the new stars with ages typically younger than
100 Myr in the secondary bar,
which are the youngest population
of the interacting galaxy,
have more top-heavy IMF.
This misalignment between the major axis of the IMF distribution
and that of the stellar bar is an unexpected result in the present study
and thus worth a further investigation. We will discuss this misalignment
more extensively in our future papers by using more models of galaxy
interaction.

\begin{figure*}
\psfig{file=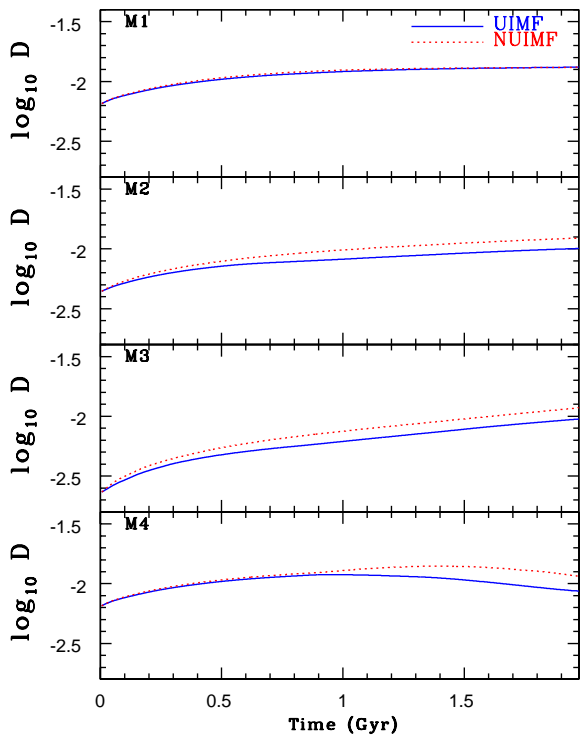,width=8.0cm}
\caption{
The same as Fig. 6 but for $D$.
}
\label{Figure. 9}
\end{figure*}

\subsection{IMF self-regulation mechanism}

SNIa and SNII feedback effects are demonstrated to prevent the high-mass
end of the IMF to become too top-heavy ($\alpha_3 <1.6$) in BM13.
By using the present new model, we can confirm or rule out such a 
suppression effect of SNe.  Fig. 5 shows the histograms of the four
IMF slopes for the three MW models, M1
(fiducial), M2 (gas-rich) , M4 (tidal interaction),
and M6 (low-mass dwarf)  with SN feedback  effects
(referred to as `SN' for convenience)
or without (`NSN').
Although the differences in the $\alpha_1$ and $\alpha_2$
histograms between the SN and NSN cases are not so remarkable
in the gas-poor model M1 (isolated) and M4 (interaction),
the $\alpha_3$ histograms are clearly different between the two.
For two MW models, a larger number of new stars can have top-heavy
IMFs with $\alpha_3 <2$ in the NSN case. 
This more top-heavy IMFs for NSN
can be more clearly seen in the gas-rich MW model M3:
the location of the peaks in the $\alpha_3$ distribution is significantly
different between the model M3 with SN and NSN. 

These results therefore confirm
that SN feedback effects can prevent  {\it the high-mass end}  of  the
IMF to become too top heavy,
because they can suppress the formation of high-density ${\rm H_2}$ regions
in galaxies. 
Here it should be noted that the suppression of
the formation of  high-density gas regions
by SN feedback effects have been already pointed by a number of recent works
(e.g., Bournaud et al. 2010; Shetty \& Ostriker 2912, Hopkins et al. 2012;
Lagos et al. 2012).
It is worth mentioning that $\alpha_1$ (and $\alpha_2$)
is  systematically flatter
in the gas-rich MW model M3 with SN 
in comparison with M3 with NSN.  This is mainly
because chemical enrichment, which can steepen the low-mass end of the IMF,
is more strongly suppressed in the model with SN.
This result implies that SN feedback effects can prevent 
{\it the low-mass end} of the IMF
to become too bottom-heavy  in star-forming galaxies with SN.
These  IMF self-regulation mechanisms found for the three MW models above
can be seen also in the low-mass dwarf model (M6).

It should be noted here that SN feedback effects can be stronger
during starbursts in the NUIMF model than in the UIMF one,
because a larger number of SNe can be
produced owing to the  more top-heavy IMF in the NUIMF model.
Therefore,  SN feedback effects in the NUIMF model
can more strongly suppress the formation of rather high-density ${\rm H_2}$ 
gas clouds from which new stars with very  top-heavy IMFs can be formed.
Thus, a variable IMF can `self-regulate' its evolution in the sense that
SN feedback effects that are controlled by the IMF can prevent
the IMF to become too top-heavy.  This self-regulation mechanism should
be very important in the chemical and dynamical evolution of starburst
galaxies, where both chemical yield from massive stars
and SN and the strength of SN feedback effects
strongly depend on the IMF slopes.

\begin{figure*}
\psfig{file=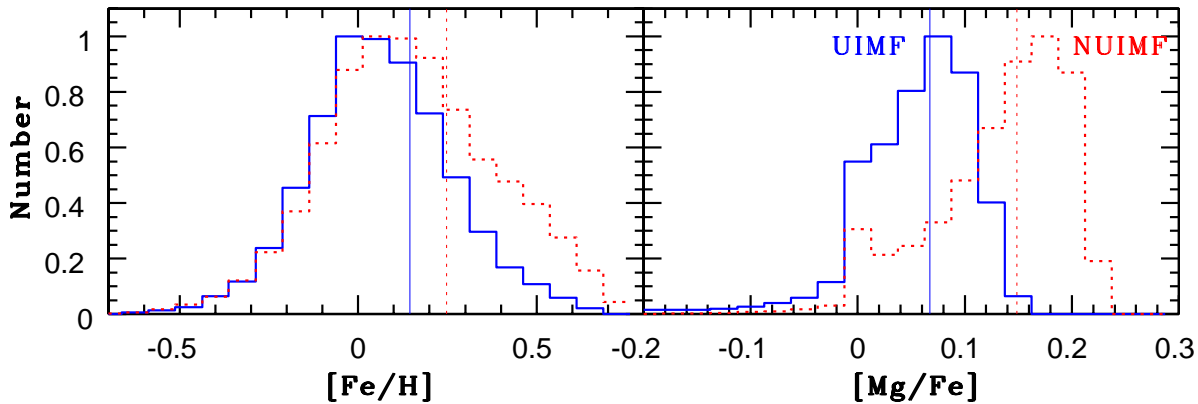,width=16.0cm}
\caption{
The metallicity distribution functions of new stars for [Fe/H] (left) and
[Mg/Fe] (right) for the UIMF (blue solid) and the NUIMF (red dotted) in the
gas-rich MW model M3. The 
`number' in the $y$-axis for these plots
is `normalized number' in the sense that the number of new stars at each
metallicity  bin 
is divided by the maximum number among all bins. 
The mean values of [Fe/H] and [Mg/Fe]
for the UIMF and the NUIMF are  indicated by blue solid and red dotted
lines, respectively.
}
\label{Figure. 10}
\end{figure*}

\section{Results}

\subsection{SFH}
The left panel Fig. 6 shows the differences in
the time evolution of SFRs between the four  MW
models (M1 - M4) with the UIMF and  the NUIMF. 
Although there are no significant  differences in
the SFR evolution between the gas-poor MW
model M1 with the UIMF and  the NUIMF,
the SFRs of 
the two gas-rich models 
(M2 with $f_{\rm g}=0.27$ and M3 with $f_{\rm g}=0.55$)
at the  early disk evolution phases ($T<0.6$ Gyr) 
can be systematically higher for the UIMF. 
This is mainly because the IMF becomes more top-heavy in the early
actively star-forming phases (`starbursts')
so that SN feedback effects can more strongly
suppress efficient star formation in the gas-rich models. 
Such stronger suppression of star formation for the NUIMF can be marginally
seen in the starburst phase ($T\sim 1$ Gyr) of the gas-poor tidal interaction
model M4. 

The right panel of Fig. 6 shows the time evolution
of the normalized gas mass ($F_{\rm g}(t)=M_{\rm g}(t)/M_{\rm g}(t=0)$)
for each of the four models.
As a result of suppressed star formation, 
gas consumption in galactic disks  can be significantly
slowed down for the NUIMF. This can be quantified by using the following normalized quantity:
\begin{equation}
R_{\rm gas}= -\frac{dF_{\rm g}(t)}{dt} = -\frac{1}{M_{\rm g}(t)} \times \frac{dM_{\rm g}(t)}{dt},
\end{equation}
where $M_{\rm g}$ is the total gas mass at each time step
and $F_{\rm g}(t)=M_{\rm g}(t)/M_{\rm g}(t=0)$. 
This normalized gas consumption rate ($R_{\rm gas}$) can be lower (i.e., gas being more slowly
consumed by star formation) in the gas-rich disks with the NUIMF. For example,
the mean $R_{\rm gas}$ ($=[F_{\rm g}(t=0)-F_{\rm g}(t={\rm 2 Gyr})]/2 {\rm Gyr}$)  
is 0.37 for the UIMF and 0.27 for the NUIMF in the model M3. This smaller $R_{\rm gas}$
can be clearly seen in the tidal interaction model with a secondary starburst (M4).
These slowed down gas consumption
is one of key effects of the NUIMF on galaxy evolution in the present study.

Fig. 7 shows the locations of star-forming regions in the four models
(M1$-$M4) on
the $\Delta {\rm SFR} - {\rm SFR}_{\rm UIMF}$ plane,
where $\Delta {\rm SFR}$ is defined as follows: 
\begin{equation}
\Delta {\rm SFR} = {\rm SFR}_{\rm NUIMF}-{\rm SFR}_{\rm UIMF}.
\end{equation}
Therefore, this figure describes 
the differences  
between the SFRs  for the UIMF (${\rm SFR}_{\rm UIMF}$)
and for the  NUIMF 
(${\rm SFR}_{\rm NUIMF}$)
as a function of ${\rm SFR}_{\rm UIMF}$
at all  time steps in the four models. 
Although there is no clear
trend in the $\Delta {\rm SFR}-{\rm SFR}_{\rm UIMF}$ relation 
for $ {\rm SFR}_{\rm UIMF}  \le 20 {\rm M}_{\odot}$ yr$^{-1}$, 
$|\Delta {\rm SFR}|$ is more likely to be larger 
for higher ${\rm SFR}_{\rm UIMF}$
for $ {\rm SFR}_{\rm UIMF}  > 20 {\rm M}_{\odot}$ yr$^{-1}$
(i.e., star formation is more strongly suppressed in more actively
star-forming regions for the NUIMF).
This implies that 
previous simulations with the UIMF might have overestimated 
SFRs of actively star-forming galaxies (if the IMF is non-universal).

\subsection{$D$ and $f_{\rm H_2}$}

Figs. 8 and 9 show the time evolution of the mean 
$D$ and $f_{\rm H_2}$ in the four
MW models for the UIMF and the NUIMF. 
The ${\rm H_2}$ fraction depends strongly on local properties
of ISM (e.g., density and radiation field) that change more rapidly
with time  whereas $D$ is an integrated property.
Therefore, the time evolution of $D$ in Fig. 9 appears to be smoother
(i.e., much less violent change with time) than that of $f_{\rm H_2}$
in Fig. 8.
Owing to the initial high gas densities
and high [Fe/H]
in the central regions of gas disks, the mean $f_{\rm H_2}$ can be higher
in the very early phase of disk evolution ($T<0.1$ Gyr).
These high-density gaseous regions  can be rapidly consumed by star formation
so that $f_{\rm H_2}$ can become smaller and start steady evolution
($T > 0.3$ Gyr). Chemical enrichment due to star formation can increase
the chemical and dust abundances of the gas disks, and consequently,
${\rm H_2}$ formation on dust grains can become more efficient.
As a result of this, $f_{\rm H_2}$ can slowly increase in the isolated
MW models (M1-3). For the tidal interaction model (M4), $f_{\rm H_2}$ can
increase significantly during strong interaction owing to the formation
of shocked high-density gaseous regions (along tidal arms).
After rapid gas consumption during tidal interaction, $f_{\rm H_2}$ can
slowly decrease.

\begin{figure*}
\psfig{file=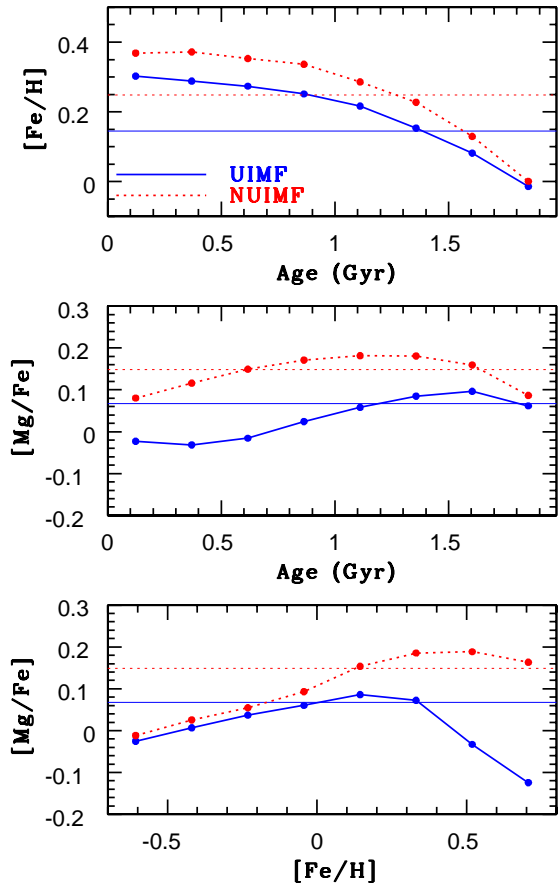,width=8.0cm}
\caption{
The age-metallicity relation of new stars  for [Fe/H] (top) and for [Mg/Fe] (middle)
and the [Mg/Fe]-[Fe/H] relation (bottom)
for the UIMF (blue solid) and the NUIMF (red dotted)  in the model M3.
The mean values of [Fe/H] and [Mg/Fe]
for the UIMF and the NUIMF  are indicated by blue solid and red dotted
lines, respectively.
The mean values of [Fe/H] and [Mg/Fe] are estimated for 8 age bins
and those of [Mg/Fe] are estimated for 8 [Fe/H] bins.
}
\label{Figure. 11}
\end{figure*}

\begin{figure*}
\psfig{file=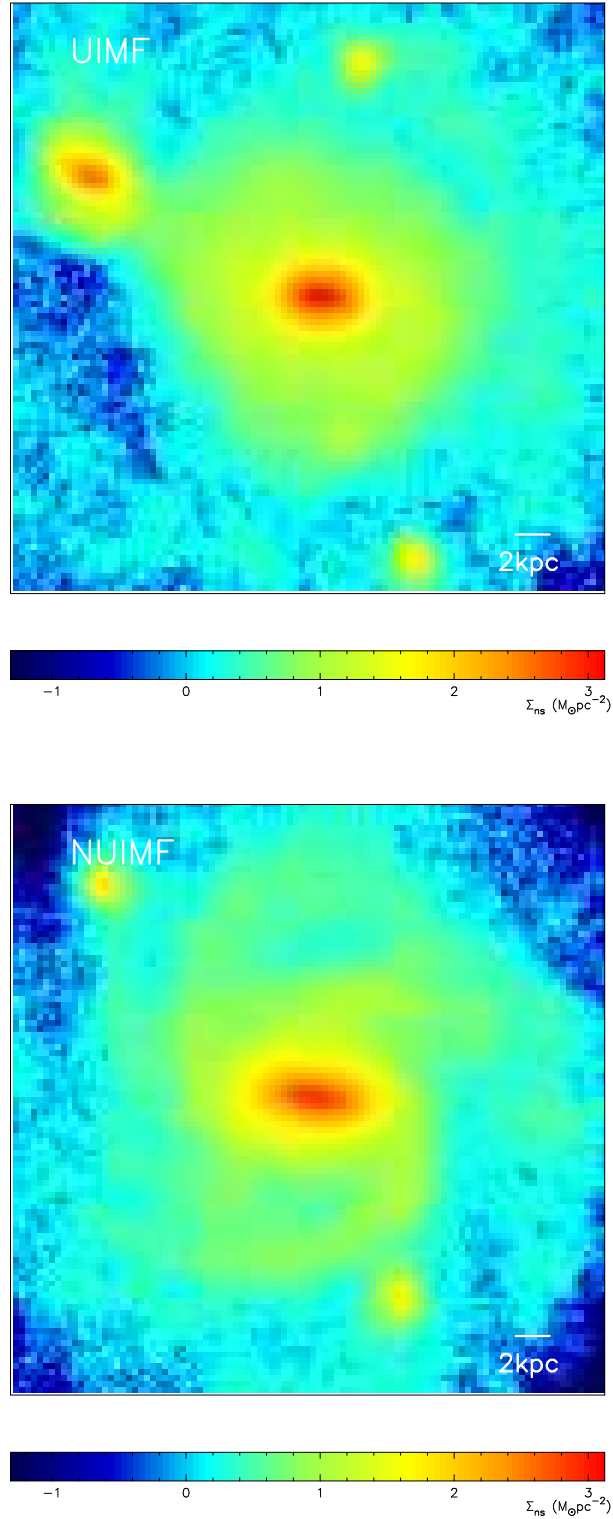,width=8.0cm}
\caption{
The 2D distribution of the projected mass densities of new stars
($\Sigma_{\rm ns}$) for the UIMF (upper) and the NUIMF (lower) in
the gas-rich MW model M3.
}
\label{Figure. 12}
\end{figure*}

\begin{figure*}
\psfig{file=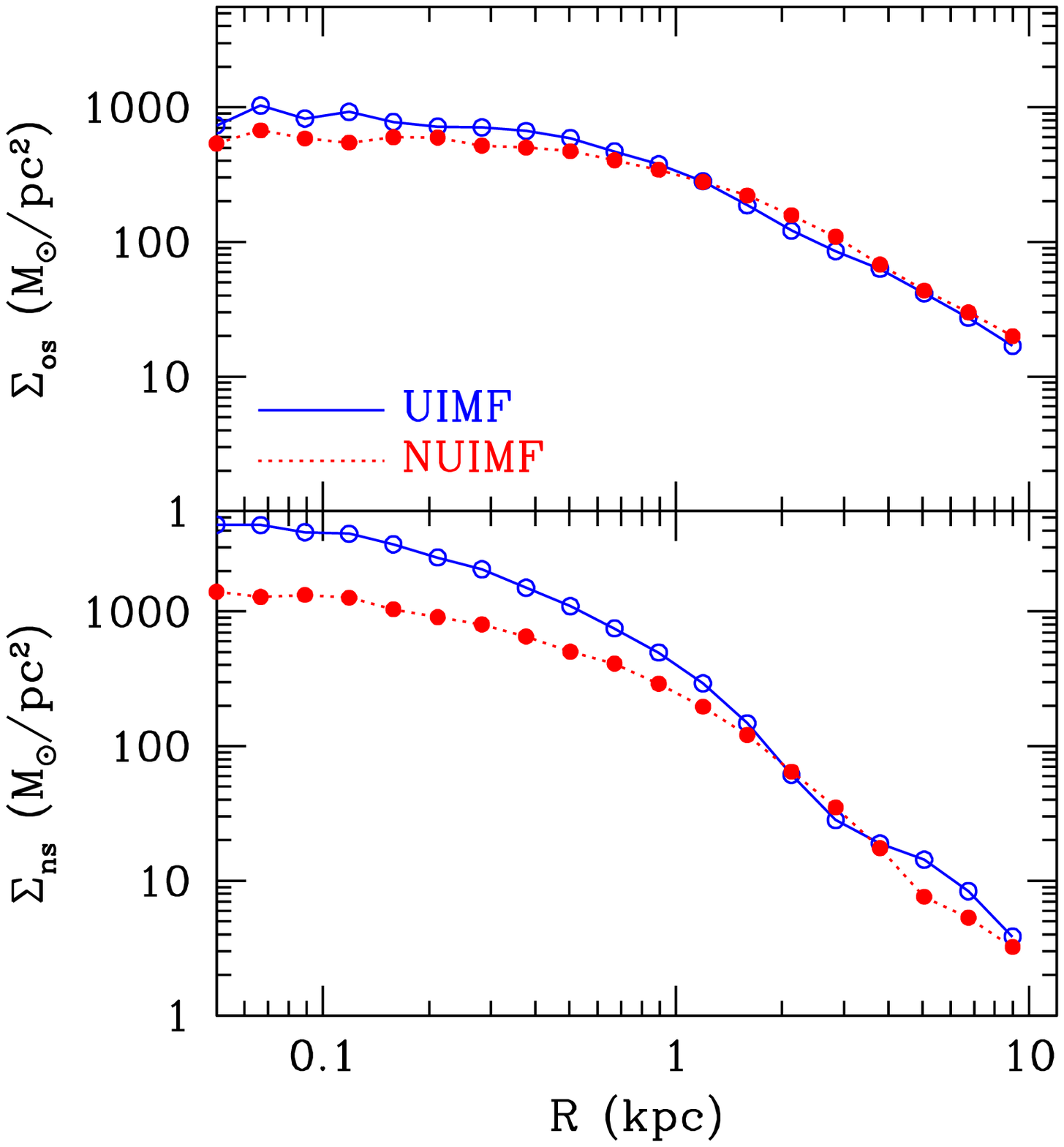,width=10.0cm}
\caption{
The projected radial density profiles of old  stars
($\Sigma_{\rm os}$, upper) and new stars ($\Sigma_{\rm ns}$, lower)
for the UIMF (blue solid, open circles) and the NUIMF (red dotted, filled circles) in
the gas-rich MW model M3.
}
\label{Figure. 13}
\end{figure*}

The evolution of $f_{\rm H_2}$ depends on the evolution of dust content
thus on whether the UIMF or the NUIMF is adopted.  Although
the differences in the $f_{\rm H_2}$ evolution between the gas-poor
model M1 with the UIMF and the NUIMF are not significant, 
$f_{\rm H_2}$ can evolve more rapidly for the NUIMF in the gas-rich models
M2 and M3. The final $f_{\rm H_2}$ can therefore become higher for the
NUIMF in the two models. The reason for these differences is that
a larger amount of dust can be produced for the NUIMF owing to the 
more top-heavy IMF in these models with more active star formation:
the larger amount of dust is responsible for more efficient ${\rm H_2}$
formation. 
Although the overall $f_{\rm H_2}$ evolution
is not so different between the UIMF and the NUIMF,
$f_{\rm H_2}$ is appreciably higher in the UIMF model during the strong
starburst phase ($T \sim 1$ Gyr). Weaker SN feedback effects in
the UIMF model are  responsible for the  higher $f_{\rm H_2}$.

The differences in $D$ evolution  between the UIMF and the NUIMF
are more remarkable in the model with higher $f_{\rm g}$. Owing to the larger
number of SNe and high-mass AGB stars formed for the NUIMF,
a larger amount of dust can be produced and mixed with ISM
in gas-rich galaxies for the NUIMF.
Consequently, $D$ evolution can proceed more rapidly
in the gas-rich models (M2 and M3) for the NUIMF. The final $D$  in these
models with the NUIMF are 
$\sim 0.1$ dex higher than those in the models with the UIMF.
Thus Figs.  8 and 9 clearly demonstrates
that $f_{\rm H_2}$ and $D$ can more rapidly increase in gas-rich,
actively star-forming disk galaxies for the models
with the NUIMF.

\subsection{Chemical properties}

The influences of the  NUIMF in chemical evolution
can be more clearly seen in gas-rich,
actively star-forming disk galaxies in the present study.
Fig. 10 shows the final [Fe/H] and [Mg/Fe] distributions
of new stars in the gas-rich model M3  for the UIMF and the NUIMF.
Although the [Fe/H] distributions are not so dramatically
different, both the mean [Mg/Fe] and the shape of the [Mg/Fe] distributions
are clearly different between the two. The model with the NUIMF
shows systematically higher [Mg/Fe] (by 0.08 dex) than the model
with the UIMF, mainly because a larger number of SN II,
which are the major contributers to the enrichment process of $\alpha$-elements
in  chemical evolution,
can be produced
in the model with the NUIMF owing to the more top-heavy IMF in the actively
star-forming regions (more precisely, the number ratio of SN II to SN Ia 
is higher for the model with  the NUIMF).

It should be stressed that the derived higher [Fe/H] and [Mg/Fe] are
not due to slightly more efficient star formation at $T>0.5$ Gyr in the NUIMF.
The gas-phase oxygen abundance 
($A_{\rm O}=12+\log {\rm (O/H)}$) is already higher 
in the NUIMF ($A_{\rm O}=9.13$) than in the UIMF (9.00) at $T=0.42$ Gyr when
star formation rate is significantly higher in the UIMF.
Furthermore, the dust-to-gas-ratio ($\log D$) is $-2.12$ for the NUIMF
and $-2.16$ for the UIMF at $T=0.42$ Gyr, which means that 
the dust abundance in ISM can  increase faster in the NUIMF even for the
early evolution phase of the disk.
This confirms
that larger chemical yields due to the more top-heavy IMF in the NUIMF
is responsible for the larger chemical abundances.
These results in M3 can be seen in both isolated and tidal interaction models
(See Appendix  A for the tidal interaction model).

Fig. 11 shows that both [Fe/H] and [Mg/Fe] of new stars are higher 
for a given age
in the gas-rich model M3 with the NUIMF owing to the more top-heavy IMF during
active star formation. It should be noted that 
[Mg/Fe] can slowly decline and thus be  kept high ($\sim 0.1$) even
$\sim 2$ Gyr after the initial active star formation phase in the model
with the NUIMF. This reflects the fact that chemical enrichment by 
SN Ia can only slowly decrease [Mg/Fe] in the model with the NUIMF.
The [Mg/Fe]-[Fe/H] relations are also clearly different between
the models with the UIMF and the NUIMF in the sense that [Mg/Fe]
for high [Fe/H] ($>0.2$)  is
remarkably higher in the model with the NUIMF.
This clear difference in the [Mg/Fe]-[Fe/H] relation can be seen in most
models in the present study.
 
As shown in Fig. 7,  star formation can be more strongly suppressed in the model M3
with the NUIMF in the early evolution phase of the disk ($T<0.5$ Gyr).
Nevertheless, the chemical enrichment can proceed more efficiently
for this model, as shown in Figs. 10 and 11. These results imply that although
the NUIMF has both positive and negative feedback effects on chemical enrichment
histories of galaxies, the positive effect (i.e., a larger amount of metals
produced) is stronger than the negative one (i.e., a smaller number of stars formed).
This has some important implications on galaxy formation, which are  discussed later
in this paper.

The above influences of the adopted NUIMF in chemical evolution of 
gas-rich galaxies can be seen even in gas-poor ($f_{\rm g}=0.09$) galaxies,
if they experience starbursts triggered by galaxy interaction. 
Fig. 12 shows that (i) the mean [Fe/H] and [Mg/Fe] of new stars 
are higher in the interaction model M4 with the NUIMF and (ii) double peaks
can be seen in the [Mg/Fe] distribution  only for the model with the NUIMF.
The peak around [Mg/Fe]$\sim 0.18$ in the model with the NUIMF
is due to the secondary starburst
triggered by tidal interaction. 
Fig. 13 shows that young stellar populations with ages less than 1 Gyr
have higher [Fe/H] and
[Mg/Fe] in the model with the NUIMF in comparison with the model with the UIMF,
as shown for the gas-rich M3 model. 
A clear difference in the [Mg/Fe]-[Fe/H]
relation for metal-rich populations with [Fe/H]$>0.4$
between the UIMF and the NUIMF can be seen in this tidal interaction model.

Radial gradients of [Fe/H] and [Mg/Fe] are significantly different
between the starburst model M4 with the UIMF and the NUIMF.
For example, the [Fe/H] gradients ($\delta {\rm [Fe/H]} / \delta R$) are
$-0.018$ dex kpc$^{-1}$ and $-0.025$ dex kpc$^{-1}$ for the UIMF
and for the NUIMF, respectively,
which means that the model with the NUIMF shows a steeper
negative [Fe/H] gradient.
The [Mg/Fe] gradients ($\delta {\rm [Mg/Fe]} / \delta R$)
in the central 12 kpc of stellar disk  are
$-0.003$ dex kpc$^{-1}$ and $-0.01$ dex kpc$^{-1}$ for the UIMF
and for the NUIMF, respectively.
These steeper negative gradients for the NUIMF
are found in other models. For example,
$\delta {\rm [Fe/H]} / \delta R$ ($\delta {\rm [Mg/Fe]} / \delta R$)
is $-0.02$ ($+3.5 \times 10^{-4}$, i.e., almost no gradient) dex kpc$^{-1}$
 for the UIMF and 
$-0.03$ ($-0.004$) dex kpc$^{-1}$ for
the NUIMF in the gas-rich MW model M3.

These results suggest that the NUIMF can play a role in the evolution of 
radial metallicity gradients of disk galaxies.
However, the derived difference in radial [Fe/H] gradients between the UIMF and the NUIMF
is not so large ($<0.01$ dex kpc$^{-1}$)
 so that we can not determine which IMF is more consistent with
the observed radial [Fe/H] gradients of galaxies (e.g., Cheng et al. 2012 for the
MW). The derived $-0.018$ and $-0.025$ dex kpc$^{-1}$ are both consistent
with the observed value ranging from $-0.013$ to $-0.066$ for different
$|z|$ (vertical distance)   for the MW (e.g., Cheng et al. 2012),
though the observed value is estimated for stars with different ages.

As shown in Figs. 11,  new  stars with high [Fe/H] ($>0.2$) are more likely
to have higher [Mg/Fe]. These new stars, which are formed with more top-heavy
IMFs during starbursts with efficient chemical enrichment,
 have larger $\alpha_1$ and $\alpha_2$ owing to their high [Fe/H]. 
These result mean that new stars with higher [Mg/Fe] formed in
starbursts in gas-rich galaxies are more likely
to have {\it bottom-heavy} IMFs for lower stellar masses 
($m_{\rm s}<0.5 {\rm M}_{\odot}$). Furthermore, new stars with higher [Fe/H]
can have more bottom-heavy IMFs for ($m_{\rm s}<0.5 {\rm M}_{\odot}$).
These results are consistent with previous and recent observations 
(e.g., Cenarro et al. 2003; Conroy \& van Dokkum 2012), though
the present study did not specifically investigate
the formation of  early-type galaxies.

\begin{figure*}
\psfig{file=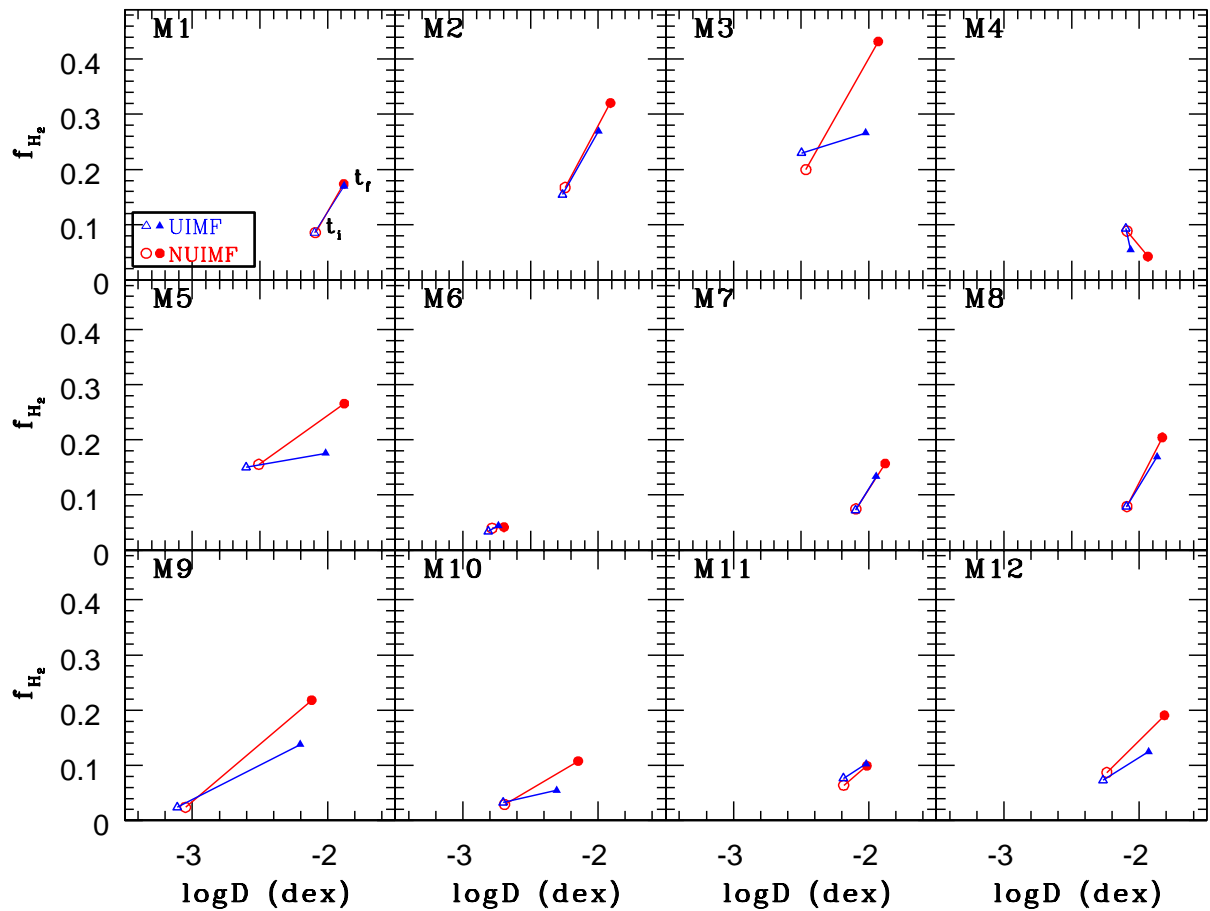,width=16.0cm}
\caption{
The locations of simulated galaxies on the $\log D - f_{\rm H_2}$ plane for
the UIMF (blue, triangle) and the NUIMF (red, circle) at $t_{\rm i}$ (open)
and $t_{\rm f}$ (filled) in the selected 12 models. The locations of a simulated
galaxy at $t_{\rm i}$ (0.14 Gyr) 
and $t_{\rm f}$ (final time step) are connected by a solid line for
the UIMF (blue) and the NUIMF (red).
}
\label{Figure. 14}
\end{figure*}

\subsection{Dynamical properties}

In the preceding subsections,  SN explosions,
the number of which
depends on the IMF,  are demonstrated to be
a key physical process that can cause possible differences in 
SFHs and chemical evolution of galaxies between models with the UIMF
and the NUIMF. It is possible that dynamical evolution of galaxies,
which can be significantly influenced by SN feedback effects,
can be different between  models with the UIMF and the NUIMF owing to the 
different numbers of SNIa and SNII between the two IMFs.
Fig. 12 shows that the distributions of new stars projected onto the $x$-$y$
plane appears to be different 
in the following three points between
the gas-rich model M3 with the UIMF
and the NUIMF. First,  more massive stellar clumps 
with higher mass densities can be formed during the disk evolution
in the model with the UIMF. Second, spiral-arm-like structures can be more 
clearly seen in the model with the NUIMF. Third, the stellar bar
within the central 2 kpc appears to be longer and thinner in the model
with the NUIMF.

Massive gaseous clumps can quickly form from local gravitational instability 
of gas disks in the gas-rich models of the present study. They can
finally become compact and bound 
stellar clumps owing to efficient star formation within
${\rm H_2}$ gas clouds. These clumps can suppress the formation
of non-axisymmetric structure such as spiral arms and bars in gas-rich disks 
and more massive clumps can have the stronger suppression effects
(e.g., Shlosman \& Noguchi 1993).
SN feedback effects during star formation
of the clumps can expel the gas in the clumps
so that it can prevent the clumps from becoming too compact and strongly
self-gravitating. The high-density gas clumps in the model with the NUIMF
can have more top-heavy IMFs, and accordingly the SN feedback effects 
can more strongly suppress the formation of compact stellar clumps.
As a result of this, less massive clumps can form and therefore less
strongly influence the formation of spiral arms and bars in the model
with the NUIMF.

More massive stellar clumps 
can more rapidly sink into the central regions of stellar disks through
dynamical friction against the disk field stars so that
they can increase the total masses of galactic bulges
and stellar nuclei (e.g., Noguchi 1999; Bekki 2010).
The more effective mass-transfer to the inner regions of disks galaxies
through clump evolution can therefore
change significantly the inner structure of the  disks.
Fig. 13 shows that the projected mass densities of new stars 
($\Sigma_{\rm ns}$) can be systematically higher for the central 1 kpc
in the gas-rich model M3
with the UIMF. This is mainly because more massive clumps
can be transferred to the inner region owing to more efficient dynamical
friction in the model with the UIMF.
Interestingly, the projected mass density of old stars ($\Sigma_{\rm os}$)
in the model with the UIMF is slightly higher, which reflects the fact
that the massive clumps transferred to the central
region contain old stars at the formation of the clumps.

Previous simulations showed that massive stellar clumps developed
in gas-rich disks can dynamically heat up the stellar disks 
and can cause the formation of thick stellar disks (e.g., Noguchi 1999).
The vertical velocity dispersion ($\sigma_{\rm z}$)
for old stars with $|z| \le 1$kpc (i.e., including stars in the thick disk)
in the model M3 
is 49.7 km s$^{-1}$ for the UIMF
and 43.1 km s$^{-1}$ for the NUIMF.
The larger $\sigma_{\rm z}$ for the UIMF is due to the more effective
scattering of old stars  by massive stellar clumps in the model with the UIMF.
This difference can be found in other models: for example, 
$\sigma_{\rm z}$ is 23.0 km s$^{-1}$ for the UIMF
and 19.4 km s$^{-1}$ for the NUIMF in the gas-rich  LMC model (M5).
These results imply  that a NUIMF can possibly influence
the time evolution of kinematical
properties of gas-rich disk galaxies.

\begin{table*}
\centering
\begin{minipage}{175mm}
\caption{Possible influences of the NUIMF on galaxy evolution}
\begin{tabular}{lll}
{Galaxy properties}
& { Physical effects of the NUIMF 
\footnote{This is based on the comparison between the models with the UIMF
and the NUIMF. For example, `higher $D$' means that $D$ is higher
in the NUIMF model than in the UIMF one.}}
& { Implications } \\
Star formation & Less enhanced SF in starburst phases of galaxies
& Slower evolution of gas content \\
Molecular hydrogen & More rapid evolution of $f_{\rm H_2}$, larger final $f_{\rm H_2}$
& Larger $f_{\rm H_2}$ in disks with higher mass densities \\
Dust &  More rapid evolution of $D$,  higher final $D$ 
& Higher  $D$ in disks with higher mass densities \\
Chemical abundances & Higher [Mg/Fe], steeper  metallicity gradients
& Higher [Mg/Fe] in poststarburst populations of E+As \\
Dynamical properties & Stronger suppression of clump formation
& Prevention of bulge growth via clump merging \\
\end{tabular}
\end{minipage}
\end{table*}

\subsection{Parameter dependences}

The fundamental influences of the adopted NUIMF on galaxy evolution
are briefly summarized as follows: (i) suppressing SFRs, (ii) accelerating
$f_{\rm H_2}$ and $D$ evolution, (iii) increasing [Fe/H] and [Mg/Fe] more
rapidly,  and  (iv) preventing the formation of massive stellar clumps. 
These four influences can be furthermore closely related to the evolution
of the distribution of SF regions, age-metallicity relations, 
radial density profiles, and stellar kinematics in disk galaxies.
This  galaxy evolution influenced by the NUIMF is clearly
shown for all models and 
briefly summarized with some implications of the results
in Table 3. The differences in the evolution rates of physical properties between 
the simulated galaxies
with the UIMF and the NUIMF are briefly discussed in Appendix A. 

The differences in the evolution of $f_{\rm H_2}$ and $D$ between galaxies
with the UIMF and the NUIMF are shown in Fig. 14 for each of selected key 12 models.
In this Fig. 14,  $f_{\rm H_2}$ and $D$ at two time steps  ($t_{\rm i}$ and $t_{\rm f}$)
are connected to show more clearly their evolution. Since $f_{\rm H_2}$ rapidly changes
in the early evolution phases of disks,  $t_{\rm i}=0.14$ Gyr is chosen (when
$f_{\rm H_2}$ can start evolving steadily). $t_{\rm f}$ represents 
the final time step in each model.
As shown in preceding sections,
some interesting results are found by comparing the MW models
(M1-M4) with the UIMF and
the NUIMF, and it is confirmed that such results can be clearly seen in other models 
(M6$-$M15)
with the UIMF and the NUIMF.
Therefore,
we briefly  describe only significant results that are worth mentioning
for these models below.
For the  better construction of this paper,
figures on the results of some of these models (M6$-$M15) are shown in the Appendix A.

\subsubsection{Low-mass disks}

The influences of the adopted NUIMF on galaxy evolution
in the low-mass disk models (LMC and dwarf models, M5, M6,  M12, and M13)
are essentially the same as those found for the MW models (M1-M3).
However,  chemical evolution of gas-rich disks appear to be 
more different between the low-mass disk models with the UIMF and the NUIMF.
For example, 
the location of the  peak in the [Fe/H] distribution
for  the gas-rich LMC model M5
can be clearly different between the UIMF and the NUIMF
whereas the [Fe/H] peak location 
for the gas-rich MW model M3 is not so clear (See Appendix A).
In these low-mass disk models,  clump formation can be severely suppressed
by SN feedback effect
both for the UIMF and the NUIMF so that dynamical evolution is not
so remarkably different between the models with the UIMF and the NUIMF.

\subsubsection{Bulgeless and big bulge}

An intriguing result for the bulgeless  model M7 ($f_{\rm b}=0$) is
that the high-mass end of the IMF ($\alpha_3$) becomes more top-heavy
after the formation of a stellar bar through bar instability in the 
stellar disk. This is because a larger amount of gas can be transferred
to the central region of the disk through dynamical influences of the bar
on the gas disk so that high-density gaseous regions can be formed
in the central region. This result implies that barred galaxies can
have more top-heavy IMFs, in particular, in their central regions.
Such significant IMF evolution can not be seen
in the big bulge model M8,
where the massive bulge can stabilize the stellar disk.

\subsubsection{Higher $f_{\rm dm}$}

Models with  $f_{\rm dm}=30.4$ (M9) and 136.8 (M10) correspond
to the earlier phase of the MW formation when the disk is still growing
by gas accretion and thus have a lower total mass (i.e., larger mass
fraction of dark halo). Significant differences
in chemical evolution for $\sim 2$ Gyr 
between the MW models with the UIMF and the NUIMF are
found for  M1$-$M3. In order to confirm
whether this is true for the {\it long-term}
evolution ($\sim 6$ Gyr) of the MW models,
we investigate the early MW models M9 and M10 with the UIMF and the NUIMF
for $\sim 6$ Gyr.
It is clear that  (i)  [Fe/H] and [Mg/Fe] distributions
in the model M9 with the UIMF and the NUIMF are significantly different
between the UIMF and the NUIMF and (ii) [Mg/Fe] differences between
the two IMFs are more significant for higher [Fe/H] (See Appendix A).  These differences
between the two IMFs are found
in M10 (and other models) too,  which means a NUIMF 
can be imprinted on the metallicity distribution
functions of galaxies.
The larger value of $f_{\rm b}$ (=0.09) for the binary fraction of stars that
can become SNe Ia is adopted in the present study so that the very metal-rich
stars after long-term chemical enrichment can have lower [Mg/Fe] ($ \sim -0.2$).

\subsubsection{LSB}

A significant difference between the standard MW model M3 and the LSB model M11
is that $\alpha_3$ is systematically larger (i.e., steeper IMF)
in the LSB model. The mean $\alpha_3$ is 2.44, which is 0.07 larger than
that for the standard MW model. Furthermore,  some minor  fraction of new stars
in the LSB model show $\alpha_3 > 3$ (but less than 3.2),
which is not seen in other models.
This result implies that (i) the IMF in LSB disk galaxies can be bottom-heavy
and thus (ii) their chemical evolution can more  slowly proceed owing to the 
less amount of metals produced in massive stars.
The final $D$ and $f_{\rm H_2}$ are lower in the LSB model in comparison with
other MW models with higher mass densities. Given that the IMF is more bottom-heavy
in the LSB model,  this result means that $D$ and $f_{\rm H_2}$ are likely
to be higher in disk galaxies with higher mass-densities.

\subsubsection{Interacting galaxies}

A larger amount of gas can be transferred to the central regions of
interacting galaxies to form high-density $\rm H_2$ gas clouds
for the tidal interaction  models
 with larger $m_2$. Therefore,  $\alpha_3$ during starbursts
triggered by tidal interaction is smaller
(i.e., more top-heavy) in M14 with $m_2=3$ than in M1 with $m_2=1$ and M15
with $m_2=0.3$. There is a positive correlation between SFRs and  $\alpha_3$
during starburst in these interaction models with starbursts 
(i.e., more top-heavy in stronger starbursts).

\begin{figure*}
\psfig{file=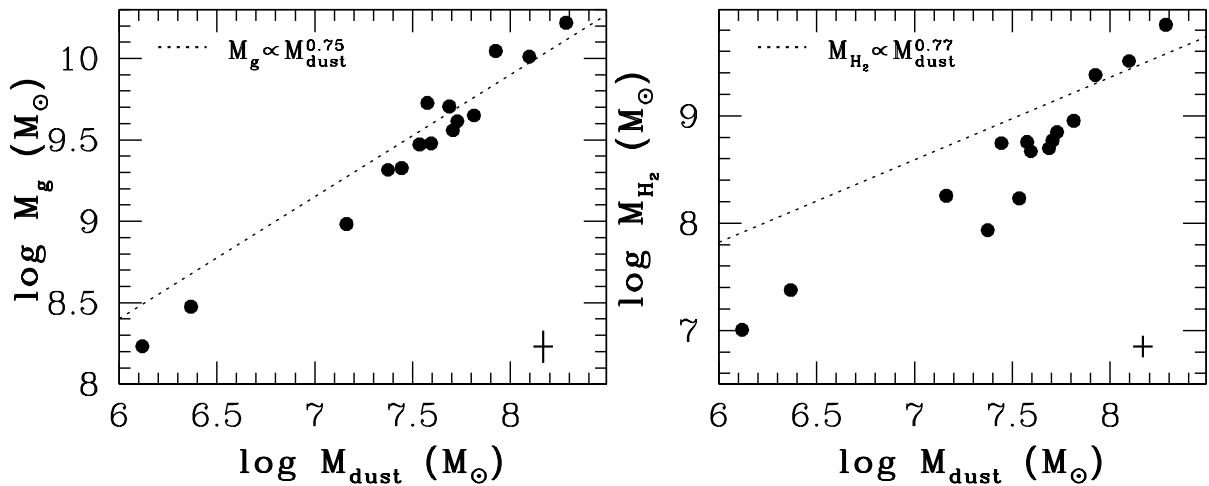,width=16.0cm}
\caption{
The plots of simulated galaxies on the $\log M_{\rm dust}-M_{\rm g}$ (left)
and $\log M_{\rm dust} - M_{\rm H_2}$ planes (right) for all 15 models with
the NUIMF. The dotted line is the observed relation by Corbelli et al. (2012).
The observational error bar is shown by a cross in each panel.
}
\label{Figure. 15}
\end{figure*}

\section{Discussion}

\subsection{Accelerated or deaccelerated galaxy evolution by the NUIMF ?}

As summarized in Table 3,  the present study has shown, for the first time,
that the time evolution of galaxy properties can be accelerated or deaccelerated 
by the NUIMF. Star formation in actively star-forming galaxies can be suppressed
by a larger number of SN explosions caused by a more top-heavy IMF 
in denser gas clouds for the NUIMF. The gas consumption  in these star-forming
galaxies can be therefore slowed down for the NUIMF.
However,  a larger amount of dust and metals can be produced by the more top-heavy
IMF so that chemical evolution can more rapidly proceed: the larger chemical yields
can have stronger influences on galactic chemical evolution
than the lower SFR for the NUIMF.  Furthermore, dust production
by SNe and AGB stars can become slightly more efficient for the more top-heavy
IMF. As a result of this, the formation efficiency of ${\rm H_2}$ can increase for
the NUIMF.

These positive and negative feedback effects of the NUIMF have not been
included in previous simulations of galaxy formation. Therefore, the formation processes
of galaxies could be significantly changed in numerical simulations of galaxy formation
based on a CDM cosmology, if the feedback effects of the NUIMF are properly included
in the simulations. A possible significant effect of the NUIMF on galaxy formation
is that final metallicities of galaxies formed with strong starbursts can be systematically
higher in galaxy formation models with the NUIMF than in those with the UIMF
(owing to the more top-heavy IMFs in starburst phases). We plan to investigate
whether and how the NUIMF can possibly influence galaxy formation in detail by
using more sophisticated simulations with realistic initial conditions of galaxy
formation based on a CDM cosmology.

\subsection{Consistency with observations}

The present simulations do not start from initial conditions of galaxy formation so that
we can not discuss whether the observed properties of galaxies
(e.g., color-magnitude relation etc)  can be reproduced by
models with the NUIMF in a self-consistent manner. For example, in order to
discuss the color-magnitude (or mass-metallicity) relation of galaxies,
we need to know initial age and metallicity
distributions of old stars, which are parameters in the present study.  
The consistency between the observed properties depending on
ages and metallicities of stars and the simulated ones will need to be done
by our future simulations of galaxy formation that can predict physical properties
of stars with different ages and metallicities.
However, we can 
discuss whether the abundance of gas and dust in simulated disk galaxies can be consistent
with the observed ones or not, because their evolution
does  not depend so strongly on the adopted assumptions of initial
ages and metallicities  of old stars.

Fig. 15 shows that the simulated correlation between total gas mass ($M_{\rm g}$)
and dust mass ($M_{\rm dust}$) is roughly consistent with the observed one by 
Corbelli et al. (2012). However, the simulated $M_{\rm dust}-M_{\rm H_2}$ relation
is slightly steeper than the observed one and it appears to  deviate significantly
from the observed one for lower $M_{\rm dust}$ ($\log M_{\rm dust}<6.5$). 
It should be noted here that observations do not have data points for 
$\log M_{\rm dust}<6.5$ (i.e., the observational line in Fig. 15 is just an extension 
of the $M_{\rm dust}-M_{\rm H_2}$ derived for galaxies with higher $M_{\rm dust}$
in Corbelli et al. (2012). This apparent deviation can be seen in our previous
models with the UIMF (Bekki 2013), which means that this is not caused by
for the NUIMF. If the observed $M_{\rm dust}-M_{\rm H_2}$ relation
($M_{\rm H_2} \propto M_{\rm dust}^{0.77}$) is confirmed for $\log M_{\rm dust}<6.5$,
then it would possibly mean that the present dust model rather than the IMF model
would need to be improved for better consistency with observations.
These results imply that the NUIMF does not have a serious problem 
(e.g., over-production of dust) in reproducing 
the observed gas and dust properties in galaxies.

\subsection{Search for chemical signatures of top-heavy IMFs in 
poststarburst galaxies}

BM13 have demonstrated that the high-mass end of the IMF ($\alpha_3$)
can become more top-heavy in interacting and merging galaxies with
strong starbursts owing to the formation of high-density gas clouds.
The present study, which incorporates a NUIMF model in galaxy-scale
chemodynamical simulations more self-consistently,  has confirmed
the result by BM13. These simulation results are consistent 
with recent observational results which have shown more top-heavy
IMFs in galaxies with clear signs of past tidal interaction
(e.g., Habergham et al. 2010).
Other theoretical studies already suggested 
a necessity of top-heavy IMFs in explaining a number of observational
properties of galaxies (e.g.,  Elmegreen 2009 for a review).
Although many observational studies tried to determine whether 
the IMF is top-heavy or not in nearby starburst galaxies,
they have not made a robust conclusion on the possible top-heavy IMF.
For example,
it has been controversial even for the nearby starburst galaxy,
M82, whether observed physical properties of M82  are consistent
with top-heavy IMFs (e.g., Rieke et al. 1980; Smith \& Gallagher 2001)
or with canonical ones (e.g.,  Devereux 1989; Satyapal 1997).

The present study has shown that [Mg/Fe] distributions
and [Mg/Fe]-[Fe/H] relations of new stars 
can be significantly different
between starburst  disk galaxies with the UIMF and the NUIMF:
[Mg/Fe] is significantly higher in the models with  the NUIMF, in particular, 
for the metal-rich stars. 
In the present models with the NUIMF, the central starburst components
can have more top-heavy IMFs and thus have higher [Mg/Fe] ($\sim 0.2$).
On the other hand,  [Mg/Fe] can be at most $\sim 0.1$ in the models
with the UIMF. The mean values  of [Mg/Fe] in starburst galaxies 
can not become so high ($\sim 0.4$), firstly because the IMF can not
become too top-heavy owing to the IMF self-regulation mechanism,
and secondly because the prompt SN Ia can more rapidly lower [Mg/Fe]
in the present models. 
These  results imply that a way to find possible
evidence of the top-heavy IMF in starbursts is to investigate [Mg/Fe]
in the central metal-rich poststarburst components in the `E+A' 
(or `K+A') galaxies. If [Mg/Fe] is as high as 0.2 in the central
region of E+As, then such a moderately 
high [Mg/Fe] is more consistently explained
by starbursts with top-heavy IMFs.

A growing number of recent spectroscopic observations of E+As 
have investigated spatial distributions of poststarburst components
in E+As.
(e.g., Pracy et al. 2005, 2012, 2013; Yagi et al. 2006;
Goto et al. 2008; Swinbank et al. 2012).
Some of these have 
revealed the strong radial gradients of Balmer absorption
lines in E+As and demonstrated that the observed gradients
are consistent with the predicted radial gradient from previous simulations
of E+A formation (e.g., Bekki et al. 2001; 2005). 
They have therefore suggested that at least some  E+As were formed as 
a result of major galaxy merging with strong central starbursts.
Given that these spectroscopic observations can investigate 
the radial gradients of [Mg/Fe] in E+As,
they will be able to confirm or rule out the present of poststarburst
populations with moderately high [Mg/Fe] ($\sim 0.2$) in the 
central regions of E+As. As shown in BM13 and the present study,
the IMF can be more top-heavy in the inner region than
in the outer ones for  starburst galaxies.
This implies that [Mg/Fe] should be higher in the inner regions of
E+As, which should be equally possible for future spectroscopic observations
to detect. 

\subsection{Bulge formation influenced by a NUIMF}

The present study has clearly shown that dynamical evolution of gas-rich disk
galaxies through the formation of massive gaseous and stellar clumps
can be influenced by a NUIMF. This is mainly because more effective
SN feedback effects for a NUIMF can suppress more severely the formation
of high-density clumps that can be transferred to the central regions
of disk galaxies through dynamical friction. This result has the following
implications on the formation of galactic bulges, given that many previous
theoretical and observational studies investigated the formation of
galactic bulges through merging of massive clumps in the central
regions of galaxies (e.g., Noguchi 1999; Elmegreen et al. 2008, 2009;
Inoue \& Saitoh 2012).
First, previous simulations using a  SN feedback model with
a {\it fixed} 
canonical IMF can significantly
overestimate the total masses of the simulated bulges 
formed as a result of clump merging, because they can underestimate the
suppression of clump formation by SN feedback effects.

Second, such previous simulations underestimate 
the mean [Fe/H] and [Mg/Fe] in the metallicity distribution functions 
of the simulated bulges. This might be true for galaxy bulges  formed
from major merging in previous simulations with a fixed canonical IMF.
Third, disk thickening by scattering of field stars by infalling
clumps can be overestimated  in
previous simulations with a fixed canonical IMF.
These strongly suggest that the formation of galactic bulge by clump
merging need to be reinvestigated by numerical simulations with a NUIMF,
if the IMF is really non-universal.
Bulge formation through clump merging/accretion is one of key issues
in secular evolution of disk galaxies. The present study suggests that
a NUIMF can play a role in other aspects of secular evolution
of disk galaxies (e.g., radial migration and formation and disappearance
of spiral arms).

\subsection{Advantages and disadvantages of the NUIMF model}

The  NUIMF model  adopted in BM13 and the present
study  can naturally explain
and the correlation between SFR densities  and
the IMF slope ($\alpha_3$) observed by M09 and G11 (BM13).
It also enables us to predict the IMF variation in different galaxies
and its influences on galaxy evolution in a self-consistent manner.
Furthermore it can predict how the IMF slopes
depend on chemical abundances of galaxies such as [Mg/Fe].
One of key predictions from the present chemodynamical simulations
is that $\alpha_1$ (and $\alpha_2$) can be larger 
(i.e., more bottom-heavy) for metal-rich stars with higher [Mg/Fe].
Although this prediction can be closely related to recent observational
results on the correlation of the IMF slope with [Mg/Fe] in elliptical
galaxies (e.g., Conroy \& van Dokkum 2012),
the possible  dependences of the low-mass end of the IMF on [Mg/Fe] will be
more extensively discussed
in the context of elliptical galaxy formation 
in our future papers.

However, it has the following three possible disadvantages in 
explaining the observed possible IMF variation in 
galaxies.
First, the present chemodynamical simulations can not show star-forming
regions with $\alpha_3 > 3.5$, which is observationally suggested
for galaxies with low surface mass densities  in M09 (see Fig. 10 in M09).
This is mainly because star formation can not occur 
where mean $\rm H_2$ densities of ISM  are low  (thus $\alpha_3$ can possibly
become large)
in the present SF model
{\it with a threshold gas density for star formation}.

Second, the present models suggest that low-mass dwarf galaxies can have
more top-heavy IMFs owing to their lower [Fe/H],
which appears to be inconsistent with recent chemical
evolution models for dwarfs in the Local Group 
(e.g., Tsujimoto 2011) which can  explain the observed abundance patterns
for a canonical IMF with a truncated upper-mass limit (i.e., less top-heavy).
Third,  physical origins for the dependences of the three IMF slopes
on [Fe/H] and $\rm H_2$ properties are not so clearly understood,
though the adopted NUIMF model (M12) is constructed by using observational
data sets that can give constraints on the IMF dependences.

These three possible disadvantages of the NUIMF adopted in the present study 
suggest that 
our  future chemodynamical models 
need to (i) adopt more sophisticated SF recipes to explain the IMF slopes
for star-forming regions in LSBs and (ii)
discuss whether or not dwarf galaxy formation with the  NUIMF 
can really  explain the observed
abundance patterns of dwarfs (it could be possible that the observations
can be explained even by the NUIMF). 
Also it is our future study to understand whether and why the three IMF
slopes (rather than the characteristic mass of an IMF with a fixed slope)
can change according to physical properties of star-forming clouds.

\section{Conclusions}

We have investigated self-consistently (i) the time evolution
of the three IMF slopes ($\alpha_1$, $\alpha_2$, and $\alpha_3$)
of the Kroupa IMF and (ii) the influences of the IMF evolution
on galaxy evolution using our new chemodynamical simulations
with a universal Kroupa IMF (UIMF)
and a non-universal Kroupa IMF(NUIMF). We have adopted the NUIMF model
by M12 and thereby investigated the differences in galaxy evolution
between the UIMF and the NUIMF. The principle results are as follows:\\

(1) The time evolution of SFRs in galaxies can be significantly different
between the present chemodynamical models
of disk galaxies with the  UIMF and the NUIMF. For example, SFRs in gas-rich
disks ($f_{\rm g} > 0.3$) can be lower in the models with the NUIMF  than in 
those with the UIMF and the differences ($\Delta$SFR) are roughly proportional
to SFRs. In the models with the NUIMF,  the high-mass end of the IMF 
can become more top-heavy (i.e., $\alpha_3$ becomes smaller) owing to
the development of high-density molecular gas clouds so that 
SN feedback effects can be stronger.
As a result of this,  SFRs can be more severely suppressed by
SN feedback effects  in the models with the NUIMF. \\

(2) Chemical evolution in actively star-forming
disk galaxies proceeds more rapidly in the models with the NUIMF than in
those with
the UIMF one, mainly because a larger amount of metals and dust can be produced
in the models with  the NUIMF. 
As a results of this, the final [Fe/H] and [Mg/Fe]
can be by $\sim 0.1$ dex higher in the models with the NUIMF, and
[Mg/Fe] for a given metallicity can be also higher in the models with the NUIMF. 
Radial metallicity gradients of [Fe/H] and [Mg/Fe] are steeper in
the models with NUIMF.
Also metallicity distribution functions (MDFs), 
age-metallicity relations (AMRs),
and [Mg/Fe]-[Fe/H] relations are different between the models
with the UIMF and the NUIMF.
New stars formed during active star formation with significant chemical
enrichment can have higher [Mg/Fe] and steeper $\alpha_1$ and $\alpha_2$,
which means that stars with higher [Mg/Fe] can have bottom-heavy IMFs. \\

(3) The time evolution of $f_{\rm H_2}$ can be slightly different between
the  models with the UIMF and the NUIMF  in that $f_{\rm H_2}$ in gas-rich disks 
($f_{\rm g} > 0.3$)  can be higher in the models with the NUIMF. This is mainly
because dust production, which is a key factor for ${\rm H_2}$ formation
in the present simulations, is more efficient in the models with the NUIMF.
The evolution of $D$ is more rapid in the models with the NUIMF so that 
the final $D$ in star-forming disks 
can be appreciably higher in the NUIMF model.
Thus, although 
$D$ and $f_{\rm H_2}$ can be slightly higher in the gas-rich models with the NUIMF,
SFR can be lower in the models owing to the stronger SN feedback effects.\\

(4) Formation of massive stellar clumps in gas-rich
disks ($f_{\rm g} \sim 0.5$)  can be more  severely
suppressed by SN feedback effects in the models with the
NUIMF model, because a larger number
of SN Ia and II can be produced in the models.  As a result of this,
massive stellar clumps can less strongly influence the dynamical evolution of 
disks. The vertical velocity dispersions of old stellar disks can be 
(by a factor of $\sim 15$\%) lower
in the models with the NUIMF  owing to weaker dynamical heating of the disks by massive
stellar clumps. The total masses of new stars in the central 1 kpc of disks
can be larger in the models with the NUIMF. \\

(5) The projected radial density profiles of new stars in the central
1 kpc of gas-rich ($f_{\rm g}$) disk galaxies can be shallower  
in the models with the NUIMF. This is mainly because inward transfer of gas
through (i) infall of massive clumps due to dynamical friction and
(ii) dynamical action of stellar bars and spirals is more severely
suppressed by SN feedback effects in the models with the NUIMF. Given that
previous bulge formation models through infall of massive stellar clumps
assumed a standard UIMF,  the present results imply that the previous
models might have overestimated the bulge growth through infall of massive
stellar clumps. \\

(6) The high-mass end of the IMF ($\alpha_3$) can become smaller (i.e., more 
top-heavy) in actively star-forming gas-rich ($f_{\rm g} \sim 0.3-0.5$) disk galaxies
and interacting galaxies in the models with the NUIMF.
However, the IMF can not become
too top-heavy ($\alpha_3 < 1.5$), because SN feedback effects become
more efficient during the top-IMF phase so that the formation of 
very high-density molecular clouds, for which $\alpha$ can be rather small,
can be more severely suppressed. 
This IMF self-regulation mechanism (or `IMF feedback')
is very important for the evolution of starburst galaxies
and  might make the IMFs of galaxies less variable. \\

(7) The present study predicts that IMF slopes can be different
between galaxies with different physical properties
(e.g., gas mass fractions and total masses) and between different
local star-forming regions in a same galaxy.
For example,  barred disk galaxies are more likely to have more top-heavy
IMFs in the central bar regions in comparison with non-barred ones.
Also, interacting disk galaxies are likely to have more top-heavy IMFs
in their central regions in comparison with isolated disk galaxies. 
We will  discuss the IMF dependences on galaxy properties
more  extensively in our future papers. \\

We have mainly discussed 
the possible influences of a NUIMF  on the evolution
of SFRs, structural and kinematical properties,
chemical abundances,  $f_{\rm H_2}$, and $D$ in star-forming disk
galaxies with or without galaxy interaction.
However, we have not clearly demonstrated that only the NUIMF model can 
explain some of the observed properties of galaxies. In our forthcoming
papers, we plan to demonstrate which physical properties of galaxies
can be better reproduced by the NUIMF than by the UIMF.

\section{Acknowledgment}
I (Kenji Bekki; KB) am   grateful to the referee  for  constructive and
useful comments that improved this paper.
Numerical simulations  reported here were carried out
on the three GPU clusters,  Pleiades, Fornax,and gSTAR 
kindly made available by International Centre
for radio astronomy research
(ICRAR) at  The University of Western Australia,
iVEC,  and the Center for Astrophysics and Supercomputing
in the Swinburne University, respectively.
This research was supported by resources awarded 
under the Astronomy Australia Ltd's
ASTAC scheme on Swinburne with support from the Australian government. 
gSTAR
is funded by Swinburne and the Australian Government's
Education Investment Fund.
KB acknowledges the financial support of the Australian Research Council
throughout the course of this work.

\appendix

\section{Some results of selected  models and evolution rates
of all models}

In this Appendix A, we describe some interesting results of selected models
and discuss the `evolution rate' of $D$, $f_{\rm H_2}$, and gas contents in 
the models with the UIMF and NUIMF.

\begin{figure}
\psfig{file=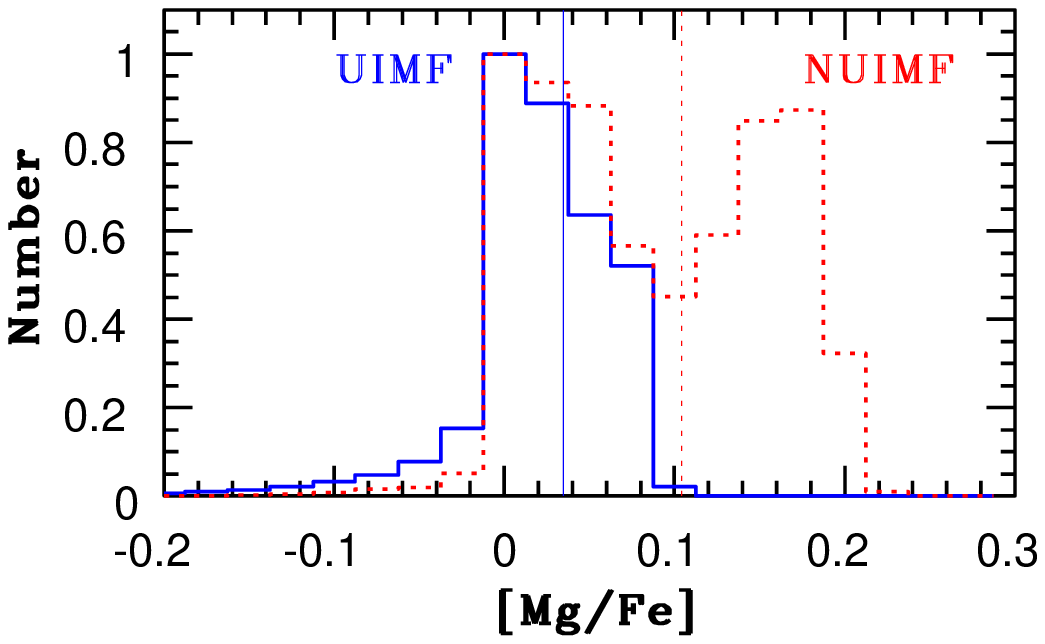,width=8.0cm}
\caption{
The [Mg/Fe] distribution of new stars in the tidal interaction model M4 for
the UIMF (blue solid) and the NUIMF (red dotted).
}
\label{Figure. 16}
\end{figure}

\begin{figure}
\psfig{file=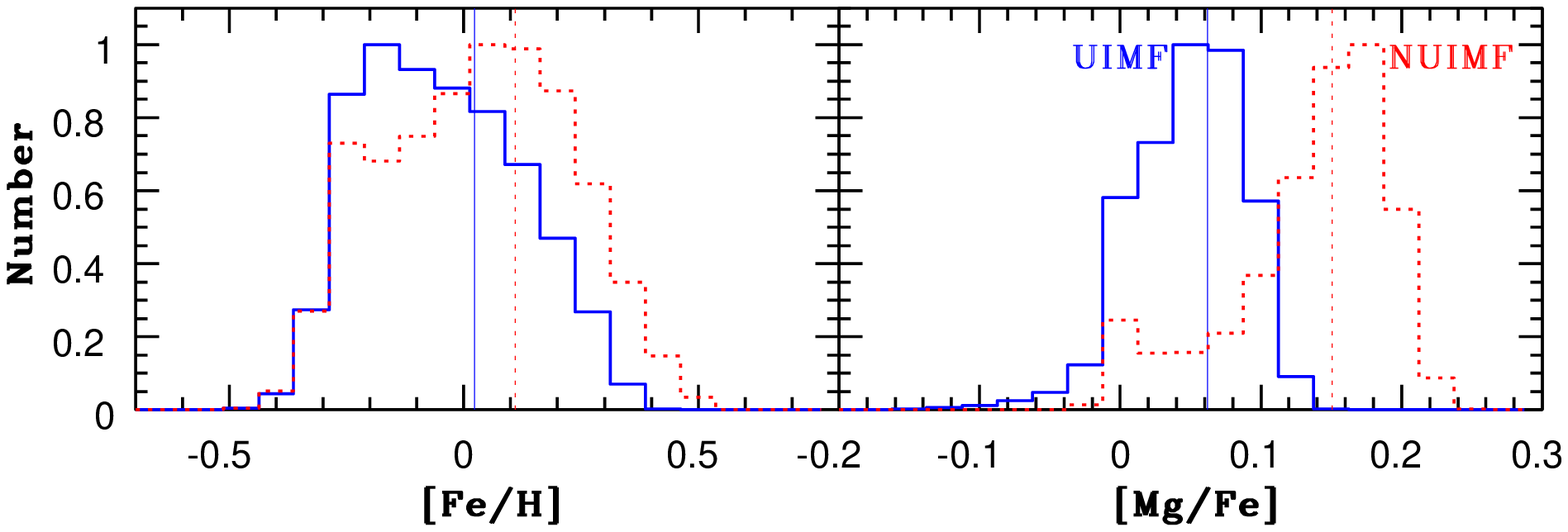,width=8.0cm}
\caption{
The same as Fig. 10 but for the gas-rich LMC model M5.
}
\label{Figure. 17}
\end{figure}

\begin{figure}
\psfig{file=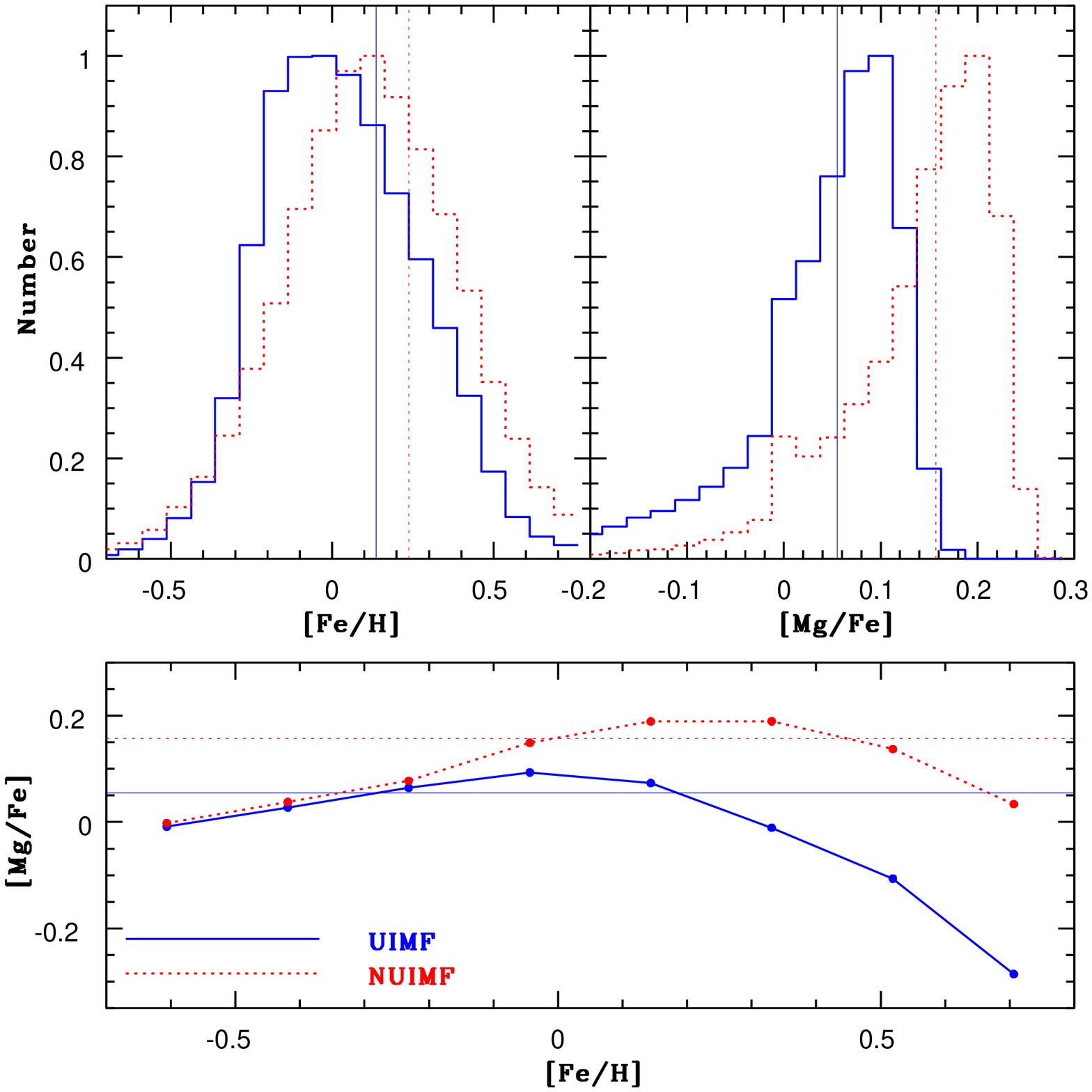,width=8.0cm}
\caption{
The [Fe/H] and [Mg/Fe] distributions (upper) and the [Mg/Fe]-[Fe/H]
relation (lower) for new stars in the MW model M10 ($f_{\rm dm}=30.4$)
with the UIMF (blue solid) and the NUIMF (red dotted). The mean [Fe/H] and [Mg/Fe]
are plotted by vertical and horizontal dotted lines in these frames.
}
\label{Figure. 18}
\end{figure}

\begin{table}
\centering
\begin{minipage}{80mm}
\caption{ The mean evolution rates of $D$, $f_{\rm H_2}$, gas-phase abundance
($A_{\rm O}=12+\log(O/H)$), and gas consumption ($F_{\rm g}$)
for the UIMF and the NUIMF. The larger values mean more rapid evolution in this table. }
\begin{tabular}{ccc}
{Physical quantity}
&  {UIMF}
& {NUIMF} \\
$d\log D/dt$  (dex Gyr$^{-1}$) & $2.6 \times 10^{-1}$ & $2.9 \times 10^{-1}$  \\
$df_{\rm H_2}/dt$  (Gyr$^{-1}$) & $2.5 \times 10^{-2}$  & $4.6 \times 10^{-2}$   \\
$-dF_{\rm g}/dt$  (Gyr$^{-1}$) & $2.2\times 10^{-1}$  & $1.8 \times 10^{-1}$  \\
$d\log A_{\rm O}/dt$  (dex Gyr$^{-1}$) & $2.2 \times 10^{-1}$  & $3.2 \times 10^{-1}$  \\
\end{tabular}
\end{minipage}
\end{table}

\subsection {M4}

The influences of the adopted NUIMF in chemical evolution of
gas-rich galaxies derived in the gas-rich MW models (M2 and M3)
can be seen even in gas-poor ($f_{\rm g}=0.09$) galaxies,
if they experience starbursts triggered by galaxy interaction.
Fig. A1 shows that 
double peaks
can be seen in the [Mg/Fe] distribution  only for the model with the NUIMF.
The peak around [Mg/Fe]$\sim 0.18$ in the model with the NUIMF
is due to the secondary starburst
triggered by tidal interaction.

\subsection {M5}

The influences of the NUIMF on the final [Fe/H] and [Mg/Fe] distributions
derived in the MW models can be clearly seen in the low-mass disk models.
Fig. A2 clearly shows that the mean [Fe/H] and [Mg/Fe] can be higher in 
the gas-rich LMC model (M5). These higher [Fe/H] and [Mg/Fe] for the NUIMF can be seen in other
low-mass disk models of the present study.

\subsection {M10}

It is confirmed that the influences of the NUIMF of the 
final [Fe/H] and [Mg/Fe] distributions and age-metallicity relations does not depend
on the baryonic mass fractions of galaxies. For example, Fig. A3 for the
MW model with a smaller baryonic fraction (M10) shows that 
(i) the final [Fe/H] and [Mg/Fe]  are higher for the NUIMF
and (ii) [Mg/Fe] for a given [Fe/H] is higher for [Fe/H]$>0$. These results imply
that the NUIMF is important for the very early chemical evolution
of galactic disks.

\subsection {Evolution rate}

In Fig. A4, the evolution rate of a physical quantity $X$
(e.g., $D$ and $f_{\rm H_2}$) is defined as follows:
\begin{equation}
\frac{dX}{dt}= \frac{ (X(t=t_{\rm f}) - X(t=t_{\rm i})  ) }  { (t_{\rm f} - t_{\rm i}) },
\end{equation}
where $t_{\rm i}$ and $t_{\rm f}$ are time at the initial and final time step
used for the $dX/dt$ estimation of a simulation,
respectively. The mean evolution rate for each physical quantity is estimated by
making an average for 15 models and the value is listed both for the UIMF and the NUIMF
in Table A1.  This table clearly shows how the evolution of $D$, $f_{\rm H_2}$, and
$F_{\rm g}$ can be different between the UIMF and the NUIMF.

For $D$ and $F_{\rm g}$, the evolution rates do not depend so strongly
on $t_{\rm f}$ (e.g., whether it is 1 Gyr or 2Gyr). Therefore, $t_{\rm i}=0$ Gyr and
$t_{\rm f}=2$ Gyr are chosen
for $D$ and $F_{\rm g}$. Owing to more violent changes of $f_{\rm H_2}$ in the
early evolution phases of gas disks, $t_{\rm i}=0.14$ Gyr rather than $t_{\rm i}=0$ Gyr can
better estimate the evolution rate for $f_{\rm H_2}$. Therefore, $t_{\rm i}=0.14$ Gyr
and $t_{\rm f}=2$ Gyr are chosen for $f_{\rm H_2}$ in most models.
For the dwarf model (M6) with the UIMF and the NUIMF,
$t_{\rm i}=0.14$ Gyr and $t_{\rm f}=0.3$ Gyr, because the evolution rate
of $f_{\rm H_2}$ can be better estimated for such $t_{\rm i}$ and $t_{\rm f}$ owing to
an intriguing behavior (more violent changes with time)
in the later evolution of $f_{\rm H_2}$.
The joint evolution of $D$, $F_{\rm g}$, and $f_{\rm H_2}$
in dwarf galaxies  need to be more extensively investigated in a separate paper.

\begin{figure*}
\psfig{file=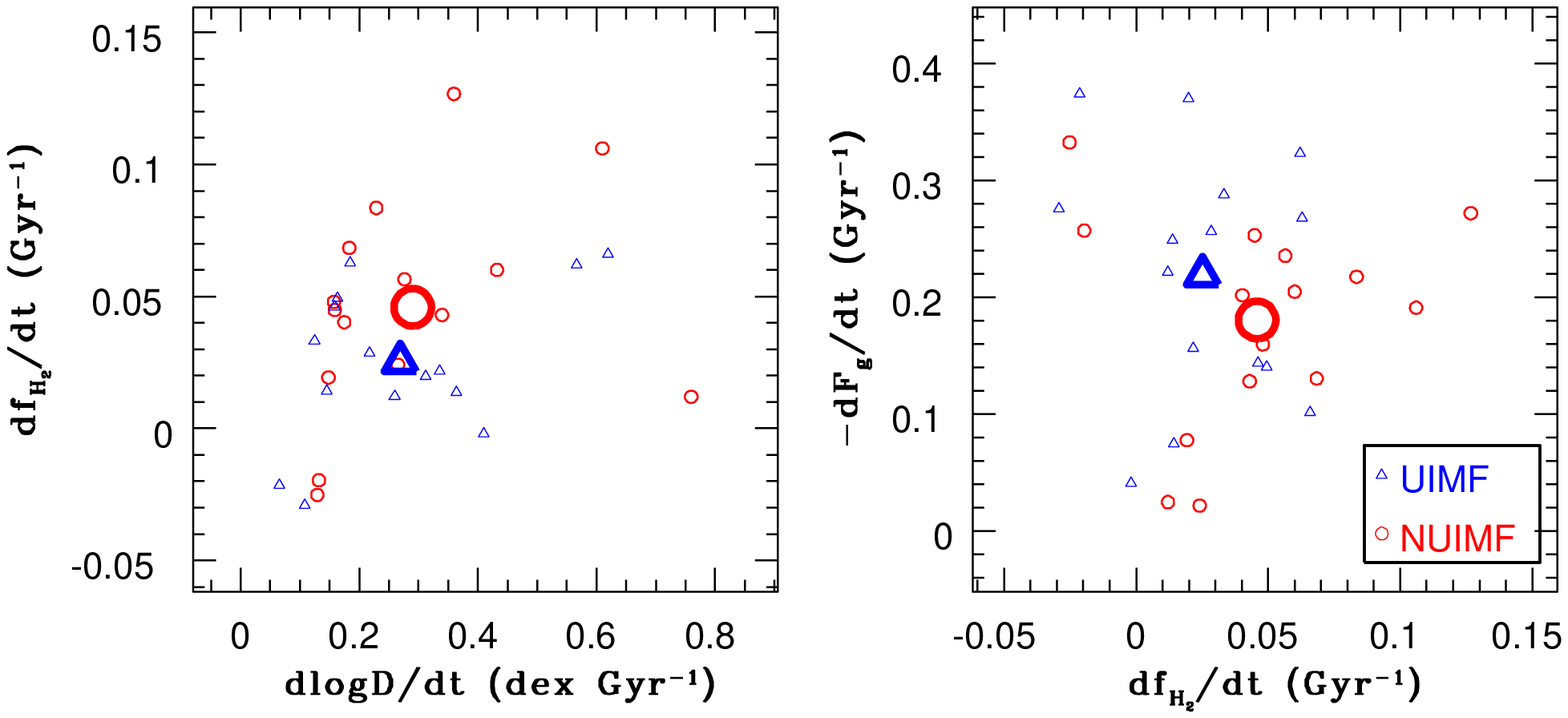,width=18.0cm}
\caption{
The plots of galaxies on the  $df_{\rm H_2}/dt - d\log D/dt$ (left)
and   $dF_{\rm g}/dt - df_{\rm H_2}/dt$ planes (right) for 
all 15 models with the UIMF (open blue triangle) and the NUIMF (open red circle).
The locations of each galaxy in these panels 
indicate the evolution rates of $D$, $f_{\rm H_2}$,
and $F_{\rm g}$. 
The big thick triangle and circle indicate the mean values of these evolution rates
for the UIMF and the NUIMF, respectively.
These evolution rates (e.g.,  $df_{\rm H_2}/dt$) depend on model parameters other than
the IMF types so that different models can show a wide ranges of the evolution rates.
These figures, however,  show that $D$ and $f_{\rm H_2}$ can increase more rapidly for the NUIMF
whereas gas can  be  consumed  more slowly by star formation for the NUIMF.
}
\label{Figure. 19}
\end{figure*}


\begin{thebibliography}{}

\bibitem[]{} 
Andrievsky, S. M., Luck, R. E., Martin, P.,  L\'epine, J. R. D., 2004, A\&A, 413, 159

\bibitem[]{} 
Bastian, N.,  Covey, K. R.,  Meyer, M. R., 2010, ARA\&A, 48, 339

\bibitem[]{} 
Baugh, C. M., et al., 2005, MNRAS, 356, 1191

\bibitem[]{} 
Bekki, K., 2009, MNRAS, 399, 2221

\bibitem[]{} 
Bekki, K., 2010, MNRAS, 401, 2753

\bibitem[]{} 
Bekki, K., 2012, MNRAS, 412, 2241

\bibitem[]{} 
Bekki, K., 2013, MNRAS,  432, 2298 (B13)

\bibitem[]{} 
Bekki, K.,  Shioya, Y.,  Couch, W.  J.,
2001, ApJL, 547, 17

\bibitem[]{} 
Bekki, K., Couch, W. J., Shioya, Y., Vazdekis, A.,
2005, MNRAS, 359, 949

\bibitem[]{} 
Bekki, K., Tsujimoto, T., 2012, ApJ, 761, 180

\bibitem[]{} 
Bekki, K., Shigeyama, T., Tsujimoto, T., 2013, MNRAS, 428, L31

\bibitem[]{} 
Bekki, K., Meurer, G. R., 2013, ApJL, 765, 22 (BM13)


\bibitem[]{}
Binney, J. \& Tremaine, S. 2007, Galactic Dynamics, 2nd ed., 
ed. J. P. Ostriker \& D. Spergel (Princeton, NJ: Princeton Univ. Press) 124

\bibitem[]{}
Bournaud, F.,  Elmegreen, B. G., Teyssier, R., Block, D. L., Puerari, I.,
2010, MNRAS, 409, 1088

\bibitem[]{} 
Cappellari, M. et al. 2012, Nature, 484, 485

\bibitem[]{} 
Cenarro, A. J.,  Gorgas, J.,  Vazdekis, A.,  Cardiel, N., \&  Peletier, R. F.
2003, MNRAS, 339, L12

\bibitem[]{} 
Cheng, J. Y., et al. 2012, ApJ, 752, 51

\bibitem[]{} 
Conroy, C., van Dokkum, P. G., 2012, ApJ, 760, 71

\bibitem[)]{}
Corbelli, E., 2012, A\&A, 542, 32

\bibitem[]{} 
Dav\'e, R.  2008, MNRAS, 385, 147

\bibitem[]{} 
Dav\'e, R., Katz, N., Oppenheimer, B. D., Kollmeier, J. A., Weinberg, D. H.,
2013, preprint (axXiv1302.3631)

\bibitem[]{} 
Decressin, T., Baumgardt, H., Charbonnel, C., \&  Kroupa, P. 2010, A\&, 516, 73

\bibitem[]{} 
Devereux, N.  A., 1989, 346, 126

\bibitem[]{} 
Duffy, A. R., Kay, S. T., Battye, R. A., Booth, C. M., Dalla Vecchia, C.,  Schaye, J.,
2012, MNRAS, 420, 2799

\bibitem[]{} 
Dutton, A. A., et al., 2011, MNRAS< 417, 1621

\bibitem[]{}
Dwek, E., 1998, ApJ, 501, 643 (D98)

\bibitem[]{} 
Elmegreen, B. G., 2004, MNRAS, 354, 367

\bibitem[]{} 
Elmegreen, B. G. 2009, in The Evolving ISM in the Milky Way and Nearby Galaxies,
Edited by  K. Sheth, A. Noriega-Crespo, J. Ingalls, and R. Paladini.

\bibitem[]{} 
Elmegreen, B. G., Bournaud, F.,  Elmegreen, D. M.,
2008, ApJ, 688, 67

\bibitem[]{} 
Elmegreen, B. G., Elmegreen, D. M.,  Fernandez, M. X.,  Lemonias, J. J.,
2009, ApJ, 692, 12
 
\bibitem[]{} 
Ferreras, I.,  La Barbera, F.,  de la Rosa, I. G.,  Vazdekis, A.,
de Carvalho, R. R., Falc\'on-Barroso, J., \&  Ricciardelli, E. 2012, MNRAS in press

\bibitem[]{}
Fu, J.,  Guo, Q.,  Kauffmann, G.,  Krumholz, M. R., 2010, MNRAS, 409, 515


\bibitem[]{} 
Geha, M., et al., 2013, ApJ, 771, 29

\bibitem[]{} 
Goto, T., Yagi, M.,  Yamauchi, C., 2008, MNRAS, 391, 700

\bibitem[]{} 
Gunawardhana, M. L. P., et al. 2011, MNRAS, 415, 1647 (G11)

\bibitem[]{}
Habergham, S. M., Anderson, J. P.,  James, P. A., 2010, ApJ, 717, 342

\bibitem[]{}
Hirashita, H., 1999, ApJ, 522, 220

\bibitem[]{}
Hopkins, P. F.; Keres, D., Murray, N., Quataert, E.,  Hernquist, L.,
2012, MNRAS, 427, 968

\bibitem[]{} 
Hoversten, E. A., \&  Glazebrook,  K. 2008, ApJ, 675, 163

\bibitem[]{} 
Inoue, S., Saitoh, T. R., 2012, MNRAS, 422, 1902

\bibitem[]{}
Lagos, C. P.,  Lacey, C. G.,  Baugh, C. M., 2012, in prepring (arXiv1210.4974)

\bibitem[]{} 
Kalberla, P. M. W., Kerp, J., 2009, ARA\&A, 47, 27


\bibitem[]{}
Kennicutt, R. C., Jr., 1998, ApJ, 498, 541

\bibitem[]{} 
Kroupa, P.,  Weidner, C.,  Pflamm-Altenburg, J.,  Thies, I.,
Dabringhausen, J.,  Marks, M., \&  Maschberger, T. 2011, preprint (arXiv1112.3340).

\bibitem[]{} 
Kroupa, P., 2001, MNRAS, 322, 231

\bibitem[]{}
Krumholz, M. R., McKee, Ch. F., Tumlinson, J.,
2009, ApJ, 699, 850



\bibitem[]{} 
Larson, R. B., 1981, MNRAS, 194, 809

\bibitem[]{} 
Larson, R. B., 1998, MNRAS, 301, 569

\bibitem[]{} 
Larson, R. B., 2005, MNRAS, 359, 211


\bibitem[]{}
Lisenfeld, U.,  Ferrara, A.,  1998, ApJ, 498, 145


\bibitem[]{} 
Marks, M., \&  Kroupa, P. 2010, MNRAS, 406, 2010

\bibitem[]{} 
Marks, M.,  Kroupa, P.,  Dabringhausen, J., \&  Pawlowski, M. S. 2012, MNRAS, 422, 2246
(M12)

\bibitem[]{}
Mckee, C. F., 1989,
in IAU Symp. 135, Interstellar Dust, Edited by Louis J. 
Allamandola and A. G. G. M. Tielens,
p431

\bibitem[]{} 
Meurer, G. R., et al. 2009, ApJ, 695, 765 (M09)

\bibitem[]{} 
Nagashima, M.,  Lacey, C. G.,  Okamoto, T.,  Baugh, C. M.,  Frenk, C. S.,  \& Cole, S.
2005, MNRAS, 363, L31

\bibitem[]{} 
Narayanan, D., \& Dav\'e, R. 2012, MNRAS, 3601, 3615

\bibitem[]{}
Navarro, J. F., Frenk, C. S.,  \& White, S. D. M. 1996, ApJ, 462, 563 (NFW)

\bibitem[]{}
Neto A. F. et al., 2007, MNRAS, 381, 1450

\bibitem[]{} 
Noguchi, M., 1999, ApJ, 514, 77


\bibitem[]{} 
Pipino, A.,  \& Matteucci, F. 2004, MNRAS, 347, 968


\bibitem[]{} 
Pracy, M. B., Couch, W. J., Blake, C.,  Bekki, K.,
Harrison, C., Colless, M.,  Kuntschner, H.,  de Propris, R.,
2005, MNRAS, 359, 1421

\bibitem[]{} 
Pracy, M. B., Owers, M. S., Couch, W. J., Kuntschner, H.,
Bekki, K., Briggs, F.,  Lah, P.,  Zwaan, M.,
2012, MNRAS, 420, 2232

\bibitem[]{} 
Pracy, M. B., Croom, S., Sadler, E., Couch, W. J., Kuntschner, H.,
Bekki, K, Owers, M. S., Zwaan, M., Turner, J., Bergmann, M.,
2013, MNRAS in press (arXiv1305.3669)


\bibitem[]{}
Rieke, G. H., Lebofsky, M. J., Thompson, R. I., Low, F. J., Tokunaga, A. T.,
1980, ApJ, 238, 24

\bibitem[]{}
Rosen, A.,  Bregman, J. N., 1995, ApJ, 440, 634

\bibitem[]{}
Shlosman, Isaac; Noguchi, M., 1993, ApJ, 414, 474

\bibitem[]{}
Smith, L. J., Gallagher, J. S., 2001, MNRAS, 326, 1027


\bibitem[]{} 
Satyapal, S., Watson, D. M., Pipher, J. L.,  Forrest, W. J.,
Greenhouse, M. A., Smith, H. A., Fischer, J., Woodward, C. E.,
1997, ApJ, 483, 148

\bibitem[]{} 
Shetty, R., Ostriker, E. C., 2012, ApJ, 754, 720

\bibitem[]{}
Sutherland, R. S., Dopita, M. A., 1993, ApJS, 88, 253

\bibitem[]{}
Swinbank, A. M., Balogh, M. L., Bower, R. G.,
Zabludoff, A. I., Lucey, J. R., McGee, S. L., Miller, C. J.,  Nichol, R. C.,
2012, MNRAS, 2012, 420, 672


\bibitem[]{}
Thornton, K., Gaudlitz, M., Janka, H.-Th.,  Steinmetz, M.,
1998, ApJ, 500, 95

\bibitem[]{} 
Treu, T., Auger, M. W., Koopmans, L. V. E., Gavazzi, R.,  Marshall, P. J.,
\&  Bolton, A. S. 2010, ApJ, 709, 1195

\bibitem[]{}
Tsujimoto, T., 2011, ApJ, 736, 113

\bibitem[]{}
Tsujimoto, T., Nomoto, K., Yoshii, Y., Hashimoto, M., Yanagida, S.,
Thielemann, F.-K.,  1995, MNRAS, 277, 945 (T95)

\bibitem[]{}
van den Hoek, L. B.; Groenewegen, M. A. T., 1997, A\&AS, 123, 305 (VG97)

\bibitem[]{} 
van Dokkum, P. G. 2008, ApJ, 674, 29

\bibitem[]{} 
van Dokkum, P. G., \& Conroy, C.  2012 ApJ, 760, 70

\bibitem[]{} 
Wolfire, M. G., McKee, C. F., Hollenbach, D., Tielens, A. G. G. M., 2003, ApJ, 587, 278


\bibitem[]{} 
Yagi, M., Goto, T.,  Hattori, T.,
2006, ApJ, 642, 152


\end{thebibliography}
\end{document}